\documentclass[12pt]{article}

\usepackage[draft]{hyperref}
\usepackage{amsmath}
\usepackage{graphicx}
\usepackage{authblk}

\usepackage{upgreek}
\usepackage{textcomp}
\usepackage{amssymb}
\usepackage{graphicx}
\usepackage{soul,booktabs,color}

\begin{document}

\title{A Review of Cavity Design for Kerr Lens Mode-Locked Solid-State Lasers}

\author{Shai Yefet and Avi Pe'er *}

\affil{Department of physics and BINA Center of nano-technology, Bar-Ilan university, Ramat-Gan 52900, Israel
	
	$^*$e-mail: avi.peer@biu.ac.il}

\date{Dated: \today}

\maketitle

\begin{abstract}
We provide a critical review of the fundamental concepts of Kerr lens \mbox{mode-locking} (KLM), along with a detailed description of the experimental considerations involved in the realization of a mode-locked oscillator. In addition, we review recent developments that overcome inherent limitations and disadvantages in standard KLM lasers. Our review is aimed mainly at readers who wish to realize/maintain such an oscillator or for those who wish to better understand this major experimental tool.
\end{abstract}


\section{Introduction}

Ultrashort pulses are key to numerous fields of science and technology. The unique temporal and spectral properties of the pulses make them an extremely valuable tool in many fields, such as the investigation of ultrafast chemical reactions \cite{femtochemistry} and multi-photon microscopy \cite{multiphotonmicroscopy}. Frequency combs are an important application of ultrashort pulses that received great attention since the Nobel Prize in Physics was awarded to J. L. Hall and T. W. H\"{a}nsch \cite{frequencycombs,Hall@nobel,Hansch@nobel}. A comprehensive review regarding the applications of ultrashort laser systems can be found in \cite{laserapplications,MLapplreview}. In the past five decades since the first demonstration of mode-locking in a He-Ne laser \cite{firstHeNemodelocking}, the field of ultrafast optics ``exploded'', and the literature is enormous. The aim of this paper cannot be, therefore, to provide a complete survey of the bibliography in the field, but rather, to focus on the experimental and practical aspects of realizing a mode-locked solid-state oscillator. Specifically, we provide a detailed description of the theoretical and experimental physics of a Kerr lens mode-locked (KLM) Ti:sapphire oscillator, oriented mainly for those interested in setting up such an oscillator or in understanding its inner workings.

The reason for choosing the Ti:sapphire oscillator as the default example in this review is trivial. The first realization of the KLM Ti:sapphire laser \cite{firstKLM}, followed by the theoretical explanation of the underlying physics \cite{piche1991,ChenWang@ML,KLM,Herrmann@abruptmodelocking}, are considered a transition point in ultrashort pulse generation. Although many gain media can produce ultrashort pulses, the Ti:sapphire crystal is by far the ultimate ``working horse'' of mode-locking, providing the shortest pulses with the highest peak powers. The excellent mechanical properties and cavity compactness of the Ti:sapphire oscillator renders it perfect for commercial production. The Ti:sapphire crystal emission spectrum in the near infra-red (NIR) regime, which spans nearly an octave in frequency, can produce ultrashort pulses down to the sub-two cycle regime \cite{shortestpulse,Sutter@shortpulse1,softaperture,Krausz@shortpulse2}. Consequently, the Ti:sapphire laser has become an extremely valuable research tool in many laboratories that focus on ultrafast phenomena and precision measurement. For these reasons, the description of the physical and experimental considerations involved in the realization of such an oscillator are beneficial to a wide audience of experimentalist (and theorists).


In this review, we collect, under a single notation, the most important topics (in our view) for an essential, but detailed, description of the KLM Ti:sapphire laser. Further details will be correspondingly referenced. In addition, special focus is given to novel cavity designs that overcome some of the inherent limitations and disadvantages of the standard design of KLM Ti:sapphire lasers, such as shaping the oscillation spectrum, lowering the mode-locking threshold and eliminating sources of astigmatism. We suggest to the reader to employ educated reading when addressing this review. While we attempted to cover all relevant topics for experimental realization of a mode-locked Ti:sapphire oscillator, many readers are probably well familiar with some of the basic concepts, such as the principles of pulsed operation or dispersion control. We therefore advise the reader to first leaf through the review and then selectively read chapters of higher interest.

\section{Outline}

Section \ref{sc.laserprincipals} provides a concise review of the basic principles of pulsed operation and mode-locking techniques. In this section, the mechanism of pulse formation is described in a qualitative and intuitive manner. In addition, a general description is given to other mode-locking techniques, which preceded Kerr lens mode-locking. Section \ref{sc:optkerreff} describes the fundamental concepts of Kerr lens mode-locking. Nonlinear effects in the spatial and temporal domains are described quantitatively. Analysis and calculation of the cavity mode in continuous-wave (CW) and mode-locked (ML) operations are reviewed in Sections \ref{sc.CW.analysis} and \ref{sc.ML.analysis}, respectively, where we describe the standard configuration of a KLM Ti:sapphire cavity and its operating regimes. Specifically, we review the calculation of linear astigmatism and methods for compensating for it and provide an analysis of the intra-cavity fundamental Gaussian mode for both CW and ML operation. Section \ref{sc:hardaperture} discusses the different types of mode-locking techniques in a Ti:sapphire oscillator: hard and soft aperture techniques are described in detail, including both quantitative and visually intuitive descriptions. Specific attention is given to the ``virtual hard aperture" method, which is less discussed in the literature, where diffraction losses in an unstable resonator produce an effective hard aperture for the laser. Section \ref{sc.additionastig} addresses the problem of nonlinear astigmatism in the cavity and how it affects the spatial mode and nonlinear response in ML operation, and the standard compensation for nonlinear astigmatism (and its disadvantages) are discussed. Dispersion management is addressed in Section \ref{sc.dispersion}: analytic expressions for calculating the group delay dispersion are provided and common dispersion compensation devices are briefly reviewed, such as prism-pairs and chirped mirrors. \mbox{Section \ref{sc.Ex.analysis}} provides a qualitative review of the common experimental behavior of a standard designed KLM Ti:sapphire cavity: the typical behavior of the mode-locking power, threshold and efficiency are described for typical and commonly used mode-locking techniques. Section \ref{sc.noveldesigns} reviews advanced cavity designs, specifically developed to improve laser performance and to overcome several inherent limitations of standard designed cavities, such as nonlinear astigmatism, Kerr lensing efficiency and intra-cavity control of the oscillation spectrum. Appendix provides further discussion regarding the compensation for linear astigmatism in its most general form.

\section{Basic Principles of Pulsed Operation in a Laser} \label{sc.laserprincipals}

In the following, we provide a survey of the basic principles of pulsed laser operation. We focus only on the few concepts that are relevant to the understanding of the Kerr lens mode-locking mechanism. For a comprehensive study of laser operation, readers are referred to the classic literature \cite{Siegman@Lasers}.

\subsection{Formation of Ultrashort Pulses}

An optical cavity can support a set of longitudinal modes that satisfy its boundary conditions: \mbox{$L=m\lambda/2$}, where $L$ is the cavity length, $\lambda$ is the mode wavelength and $m$ is an integer number. The modes that can actively oscillate in the cavity are only modes whose frequency lies within the emission spectrum of the active medium. When a laser operates in CW, multiple longitudinal modes can oscillate in the cavity simultaneously, whose relative phases are random. Therefore, the modes are randomly interfering with one another, and the laser output in time is a noisy continuous intensity, which is periodic in the cavity round trip time. A laser is said to be mode-locked if many longitudinal modes are oscillating together with a well defined phase relation between them, as opposed to CW operation, where the phases are random. This phase relation causes the modes to constructively interfere only within a short period of time, while destructively interfering at all other times, forming a pulse with a high peak intensity. \mbox{In other words}, the mode-locking mechanism contracts the electro-magnetic energy, which would otherwise spread over a long period, into an extremely short pulse, which oscillates back and forth in the cavity. Eventually, ML reaches a steady state (due to limiting mechanisms, such as dispersion, gain bandwidth, \textit{etc.}), producing a stable pulse that circulates in the resonator. Pulse formation can be qualitatively modeled based on the Kuizenga--Siegman theory \cite{FMAMmodelocking1}, describing the interplay between two mechanisms: first, pulse shortening due to some nonlinear effect, which monotonically decreases in magnitude as the pulse becomes shorter and shorter; and second, the finite gain bandwidth of the active medium, which acts as a pulse lengthening mechanism in time, which is monotonically enhanced as the pulse becomes shorter and shorter. When these two mechanisms are in balance, the pulse reaches a steady state.



\newpage

\subsection{Active/Passive Mode-Locking} \label{sc.MLtechnique}

As a preface to the Kerr lens mode-locking mechanism, we provide a general and intuitive description of the mode-locking techniques that preceded Kerr lens mode-locking. We refer the reader interested in a detailed study to the corresponding literature \cite{FMAMmodelocking1,FMAMmodelocking2,saturableabsorber,fastsaturableabsorber}, but a quick review is important to set the background for the Kerr lens mode-locking technique. In general, mode-locking establishes preference for pulsed operation, either in loss or in gain. Mode-locking techniques are either \textbf{active}, where an external modulation inside the cavity enforces pulsed operation, or \textbf{passive}, where the preference for pulsed operation is introduced by an additional intra-cavity nonlinear response.

Active mode-locking can be achieved by intra-cavity amplitude modulation (AM) or frequency modulation (FM) at an exact multiple of the cavity repetition rate, $f_{rep}$ \cite{FMAMmodelocking1,FMAMmodelocking2,harmonicML,SSB}. The first method uses the modulator as a fast shutter in the cavity, where pulsed operation is achieved by synchronizing the modulation rate with the round trip time of the pulse, so a pulse will always pass through the modulator when the ``shutter is open'', cutting away the pulse edges and pushing it towards shorter periods. The second method introduces a phase/frequency shift to light arriving outside of the temporal window of mode-locking, thus constantly pushing its spectrum, outside of the gain bandwidth. Pulses can therefore evolve only during the short time window of ``no phase'', similar to the window of \mbox{``no loss''} with \mbox{AM modulation.}

Passive mode-locking is usually achieved by introducing an effective saturable absorber into the cavity, which reduces the loss for high intensity light due to the saturation of the absorbing medium \cite{saturableabsorber,keller@MLreviewandSESAM,SESAME}. Therefore, CW will suffer constant absorption losses, whereas a pulse with a high peak power will quickly saturate the absorber by the leading part of the pulse, considerably reducing the loss for the main part of the pulse. The saturable absorber then acts as a shutter that the pulse itself activates with a much faster modulation rate than electronically-based shutters. For short pulses, one prefers a fast response of the saturable absorber to intensity variations (short recovery time) \cite{fastsaturableabsorber}.

\section{The Optical Kerr Effect} \label{sc:optkerreff}

The optical Kerr effect is the nonlinear mechanism at the core of ultrashort pulse formation in many solid-state lasers. We provide a quantitative description of the nonlinear response and discuss how it affects the laser light in both the spatial and time/frequency domains. Since both the spatial and temporal effects are nonlinear (\emph{i.e.}, intensity-dependent), they are coupled together through the master equation of pulse propagation. A more detailed and mathematical review of the master equation and pulse dynamics, laser noise, the theory of active/passive mode-locking, along with slow/fast saturable absorbers and Kerr media can be found in \cite{Ippen@MLreview,Brabec@MLreview,Haus@modelocking,MLreview@fujimoto,MLreview@brabec}.

The optical Kerr effect is a third-order nonlinear process, where the refractive index of the material is intensity-dependent \cite{kerreffect}, given by:
\begin{align}
 \label{eq:kerreffect}
 n(I)=n_{0}+n_{2}I
\end{align}
where $n_{0}$ and $n_{2}$ \cite{Kerrcoeficient} are the linear and nonlinear refractive indices, respectively, and $I$ is the light intensity. As explained below, the Kerr effect can effectively introduce an artificial saturable absorber to the cavity, placing it under the category of passive mode-locking \cite{KLMtheory}. Since the Kerr effect in transparent media is practically instantaneous, this effective saturable absorber is exceedingly fast, allowing the generation of pulses as short as a few femtoseconds, far beyond what can be achieved with ``real'' saturable absorbers (in the picosecond scale). In practice, the pulse duration of Kerr lens mode-locked lasers will be limited by the dispersion management in the cavity, the gain bandwidth of the active medium and the optical period of the carrier wave. Since $n_{2}$ for the sapphire crystal is relatively low ($\approx$$3\times10^{-16}$ cm$^{2}$/W) \cite{valueofn2}, only high peak power pulses will be sensitive to this effect.

The preference for pulses via the optical Kerr effect is rooted in its influence on the pulse profile in both space and time. In the transverse spatial domain, due to the Gaussian-shaped laser beam, the center of the beam experiences a larger refractive index compared to the wings, producing an effective focusing lens for the laser. The effective nonlinear lens can be calculated by considering the phase accumulated after propagating through a Kerr medium with thickness $z$ and a nonlinear coefficient, $n_{2}$, given by: $\phi=k_{0}n(I)z$ (to be considered as thin, the material thickness, $z$, should be considerably smaller than the Rayleigh range of the mode). By comparing the result to the well-known phase of a simple lens, the dioptric power of the nonlinear Kerr lens can be easily identified as:
\begin{align}
 \label{eq:kerrlens}
 \frac{1}{f_{NL}}=\frac{4n_{2}z}{\pi}\frac{P}{w^{4}}
\end{align}
where $P$ is the pulse peak power and $w$ is the mode radius. Note that Equation~\eqref{eq:kerrlens} holds only for a circular beam where $w_{x}=w_{y}=w$. For a non-circular beam with $w_{x}\neq w_{y}$, the dioptric power will be different for each plane, and both planes will be coupled by the nonlinear interaction. A generalized expression for the dioptric power of each plane can be found in \cite{astigmaticselffocusing}.

In combination with an intra-cavity aperture, this intensity-dependent nonlinear lens can be used as an intensity-dependent loss element that generates self amplitude modulation (SAM), which favors pulsed operation over CW operation. In general, strong SAM is desired in order to achieve robust mode-locking over a wide pump power range. Two major concepts are used when introducing an aperture into a cavity: hard and soft aperture; where hard aperture exploits a real, physical aperture in the cavity, while soft aperture exploits the pump beam size inside the laser crystal as the effective aperture. Both methods are explained in detail in Section \ref{sc:hardaperture}.

In the temporal domain, the intensity-dependent refractive index modulates the instantaneous phase of the pulse according to the instantaneous intensity by: $\phi(t)=\omega_{0}t+\phi_{NL}=\omega_{0}t+k_{0}n(I)L$; where $L$ is the thickness of the Kerr medium. This process is known as self phase modulation (SPM), which results in modulation of the instantaneous frequency around the central frequency, $\omega_{0}$, given by: $\omega(t)=\dot{\phi}=\omega_{0}+\Delta\omega(t)$. The deviation, $\Delta\omega(t)$, from the central frequency is given by:
\begin{align}
 \label{eq:SPM3}
 \Delta\omega(t)=\epsilon t\exp\left(-\frac{t^{2}}{\sigma_{t}^{2}}\right)
\end{align}
where $\epsilon=(2\omega_{0}Ln_{2}I_{0})/(c\sigma_{t}^{2})$ is the modulation coefficient and a Gaussian pulse was assumed with duration $\sigma_{t}$ (from here on, the linear refractive index, $n_{0}$, will be denoted as $n$, unless specifically mentioned otherwise). From Equation~\eqref{eq:SPM3}, one can see that the instantaneous frequency is red-shifted (to lower frequencies) in the leading edge of the pulse and blue-shifted (to higher frequencies) in the trailing edge. This process of self phase modulation (SPM) \cite{selfphasemodulation} continuously broadens the spectrum of the pulse. Thus, SAM and SPM are the mechanisms for pulse shortening and spectral broadening, respectively. Eventually, both mechanisms will be limited by dispersion compensation and the finite gain bandwidth, which leads to the formation of stable, soliton-like pulses.

\section{Cavity Analysis in CW Operation} \label{sc.CW.analysis}

In this section, we provide a detailed analysis of the optical cavity in CW: cavity stability zones, stable Gaussian modes and linear astigmatism compensation, which are critical topics for the realization of the optical cavity. A detailed description regarding the Ti:sapphire crystal properties can be found \mbox{in \cite{crystalstruct,crystalgrowth}}, including: molecular/electronic structure, absorption/emission spectra, crystal growth techniques, doping level, figure of merit, \textit{etc.}

\subsection{Basic Configuration of a Ti:sapphire Oscillator}

\begin{figure}
 \centering
 \centerline{\includegraphics[page=5, clip, trim=1.7cm 3.1cm 0cm 1.7cm, width=12cm]{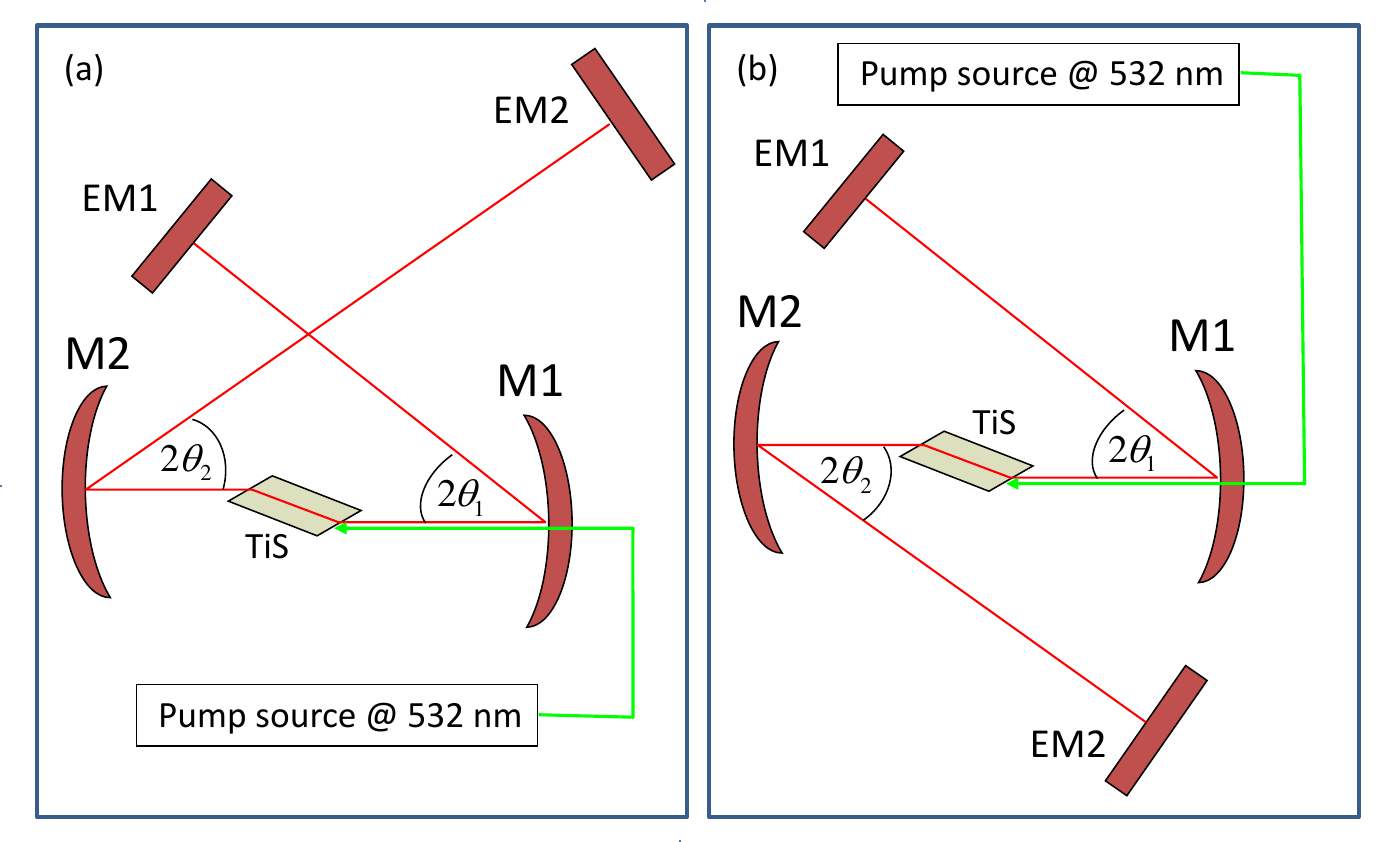}}
 \caption[Folded cavity for astigmatism compensation]
 {{Standard (\textbf{a}) X-shaped and (\textbf{b}) Z-shaped configurations of a linear Ti:sapphire cavity with a Brewster-cut crystal. Both configurations work equally well. The folding angles, $\theta_{1}$ and $\theta_{2}$, allow for the compensation of the astigmatism from the Brewster crystal, as discussed in Section~\ref{sc.linearastig}.}}
 \label{fig:foldedcavity}
\end{figure}

Figure \ref{fig:foldedcavity} illustrates the basic configuration of a linear X-shaped, or Z-shaped cavity, which are commonly used for Ti:sapphire lasers. The feedback end mirrors, $EM1$ and $EM2$, are the output-coupler (OC) and the high-reflector (HR), respectively. Two curved mirrors, $M1$ and $M2$, with foci $f_{1}$ and $f_{2}$ are used to focus the laser mode into the crystal. The distance between the curved mirrors controls the spatial mode and stability range of the cavity. The cavity has two arms: from $M1$ to $EM1$ with a length, $L_{1}$, and from $M2$ to $EM2$ with a length, $L_{2}$. The most useful technique to analyze light beams in an optical cavity is using ABCD matrices. The ABCD matrices (formulated within the paraxial approximation) are used to represent an optical system as a \mbox{2 $\times$ 2} matrix, so one can calculate how the optical system affects the path and properties of a beam of light passing through it. The beam at the system input is represented by a vector, $\mathbf{V}_{in}$, containing the distance, $x_{in}$, above the optical axis and the beam angle, $\theta_{in}$, with respect to the optical axis. The beam at the output of the system can be calculated by $\mathbf{V}_{out}=\mathbf{M}\mathbf{V}_{in}$, where $\mathbf{M}$ is the optical system ABCD matrix:
\begin{align}
 \label{eq:matrixvector}
 \left[ \begin{array}{c} x_{out} \\ \theta_{out} \end{array} \right]=
 \begin{bmatrix} A & B \\ C & D \end{bmatrix}
 \left[ \begin{array}{c} x_{in} \\ \theta_{in} \end{array} \right]
\end{align}

A detailed description of this method along with explicit expressions for the ABCD matrices of the optical elements in a cavity can be found in \cite{matrixformalism}. For the cavity illustrated in Figure~\ref{fig:foldedcavity}, we can define the distance between the curved mirrors as: $f_{1}+f_{2}+\delta$ ($\delta$ measures the shift from a perfect telescope) and use the ABCD matrix method to find the range of values for $\delta$ where the cavity is spatially stable. \mbox{The condition for} resonator stability is $|A+D|<2$, where $A$ and $D$ are the elements of an ABCD matrix that represents a single round trip in the cavity with respect to an arbitrary reference plane. Another optional cavity configuration is a ring configuration \cite{ringcavity} with the benefit of a shorter cavity length and a higher repetition rate. The same ABCD matrix technique and stability condition hold for a ring cavity.

\subsection{Stability Ranges and Limits}

When solving the stability requirements of the cavity illustrated in Figure~\ref{fig:foldedcavity}, one finds that the laser is stable for two bands of $\delta$ values bounded between four stability limits, $\delta_{0}<\delta<\delta_{1}$ and $\delta_{2}<\delta<\delta_{3}$, given by:
\begin{align}
 \label{eq:stabilitylimits}
 \delta_{0}
 & =
 0,
 &
 \delta_{1}
 & =
 f_{2}^{2}/(L_{2}-f_{2}),
 &
 \delta_{2}
 & =
 f_{1}^{2}/(L_{1}-f_{1}),
 &
 \delta_{3}
 & =
 \delta_{1}+\delta_{2}
\end{align}

Note that for simplification, Equation~\eqref{eq:stabilitylimits} was calculated without the Ti:sapphire crystal between $M1$ and $M2$. The addition of Ti:sapphire crystal (or any material with thickness $L$ and refractive index $n$) will shift all the stability limits, $\delta_{i}$, by a constant factor of $L(1-1/n)$. However, Equation~\eqref{eq:stabilitylimits} can still be used without any change by redefining the distance between $M1$ and $M2$ in Figure~\ref{fig:foldedcavity} to be: $f_{1}+f_{2}+L(1-1/n)+\delta$.

\begin{figure}
 \centering
 \centerline{\includegraphics[page=3, clip, trim=1.7cm 1.6cm 2.1cm 1.4cm, width=9cm]{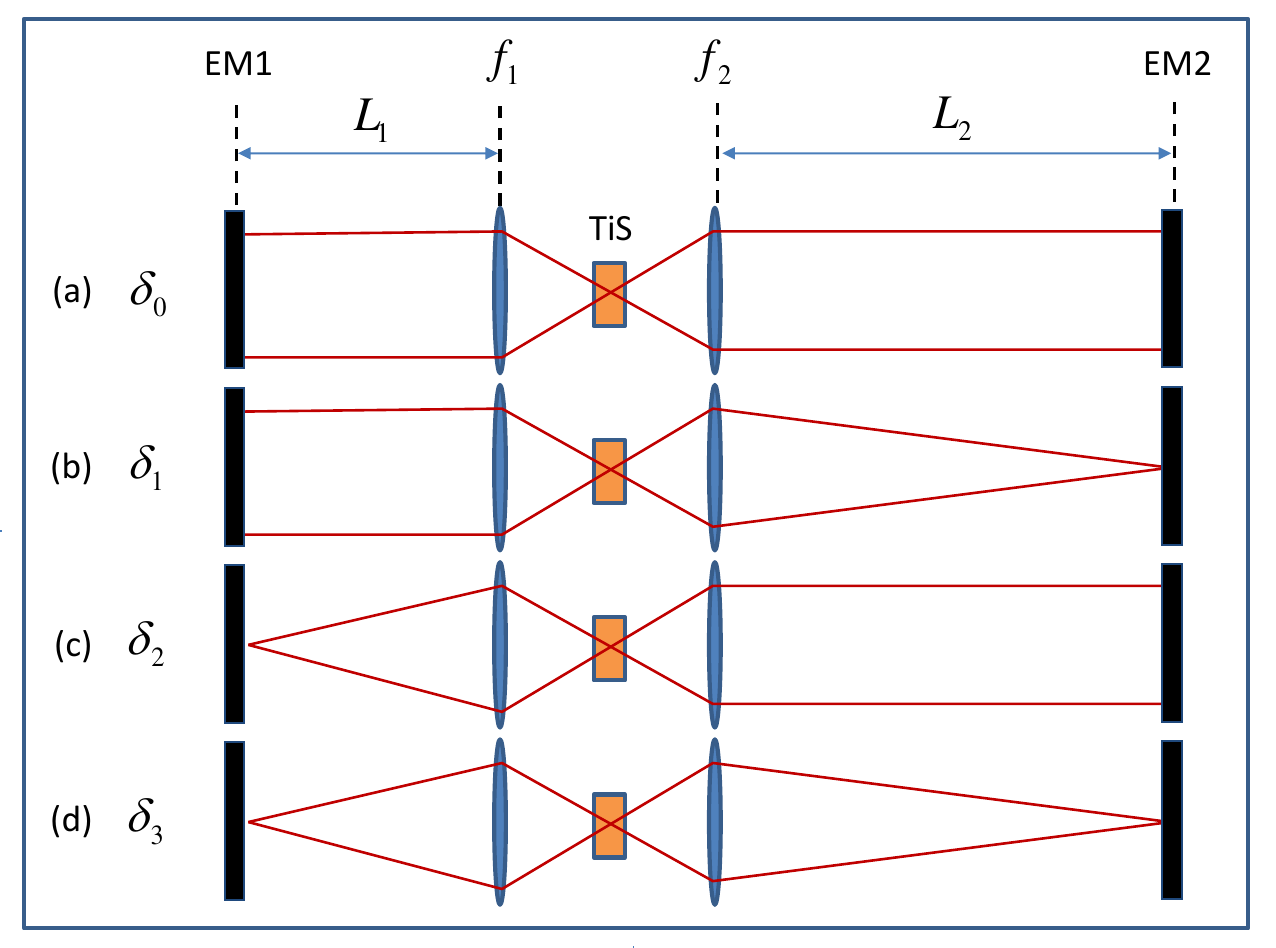}}
 \caption[Geometrical representation of the cavity modes at the stability limits]
 {{Geometrical representation of the cavity mode at the stability limits: \mbox{(\textbf{a}) plane-plane} limit; (\textbf{b}) plane-point limit; (\textbf{c}) point-plane limit; (\textbf{d}) point-point limit.}}
 \label{fig:geometricMODES}
\end{figure}

Near the stability limits given in Equation~\eqref{eq:stabilitylimits}, the stable cavity mode can be intuitively visualized using geometrical optics, as illustrated in Figure~\ref{fig:geometricMODES}. The four stability limits can be named according to the mode size behavior on the cavity end mirrors, as follows: (1) plane-plane limit (Figure~\ref{fig:geometricMODES}a); the curved mirrors (illustrated as lenses) are separated by $f_{1}+f_{2}$, forming a perfect telescope, which produces a collimated beam in both arms; (2) plane-point limit (Figure~\ref{fig:geometricMODES}b); the curved mirrors are separated by $f_{1}+f_{2}+\delta_{1}$, such that the focal point between the curved mirrors is imaged on $EM2$; (3) point-plane limit (Figure~\ref{fig:geometricMODES}c); the curved mirrors are separated by $f_{1}+f_{2}+\delta_{2}$, and the focal point between the curved mirrors is imaged on $EM1$; (4) point-point limit (Figure~\ref{fig:geometricMODES}d); the curved mirrors are separated by $f_{1}+f_{2}+\delta_{3}$, and the focus point between the curved mirrors is imaged on both the end mirrors. Kerr lens mode-locking usually occurs near one of the stability limits, as explained in Section~\ref{sc:hardaperture}.

\subsection{Gaussian Modes}

Using ABCD matrices, one can calculate the fundamental Gaussian $TEM_{00}$ mode of the cavity for CW operation inside the stability ranges of the resonator. The Gaussian mode is represented by the complex beam parameter $q=z+iz_{R}$, where $z_{R}$ is the Rayleigh range of the beam given by $z_{R}=\pi w_{0}^{2}/\lambda$ and $w_{0}$ is the mode radius at the beam focus $z=0$. To calculate the CW complex beam parameter, $q_{cw}$, at any reference plane in the cavity, one can represent the complete round trip in the cavity as an ABCD matrix with respect to a chosen reference plane. The complex beam parameter at the reference plane can be calculated by solving:
\begin{align}
 \label{eq:invqsilution}
 \left(\frac{1}{q_{cw}}\right)^{2}+\frac{A-D}{B}\left(\frac{1}{q_{cw}}\right)+\frac{1-AD}{B^{2}}=0
\end{align}
where $1/q_{cw}$ represents the beam spot size, $w$, and the beam radius of curvature, $R$, at the reference \mbox{plane by:}
\begin{align}
 \label{eq:invqdefinition}
 \frac{1}{q_{cw}}=\frac{1}{R}-i\frac{\lambda}{\pi w^{2}}
\end{align}

Solving Equation~\eqref{eq:invqsilution} provides expressions for $w$ and $R$ as a function of the ABCD matrix elements, given by:
\begin{align}
 \label{eq:sizeandradius}
 \begin{split}
 R
 & =
 \frac{2B}{D-A} \\
 w^{2}
 & =
 \frac{|B|\lambda}{\pi}\sqrt{\frac{1}{1-(A+D)^{2}/4}}
 \end{split}
\end{align}

\begin{figure}
 \centering
 \centerline{\includegraphics[page=8, clip, trim=1.9cm 1.5cm 1.9cm 2.2cm, width=9cm]{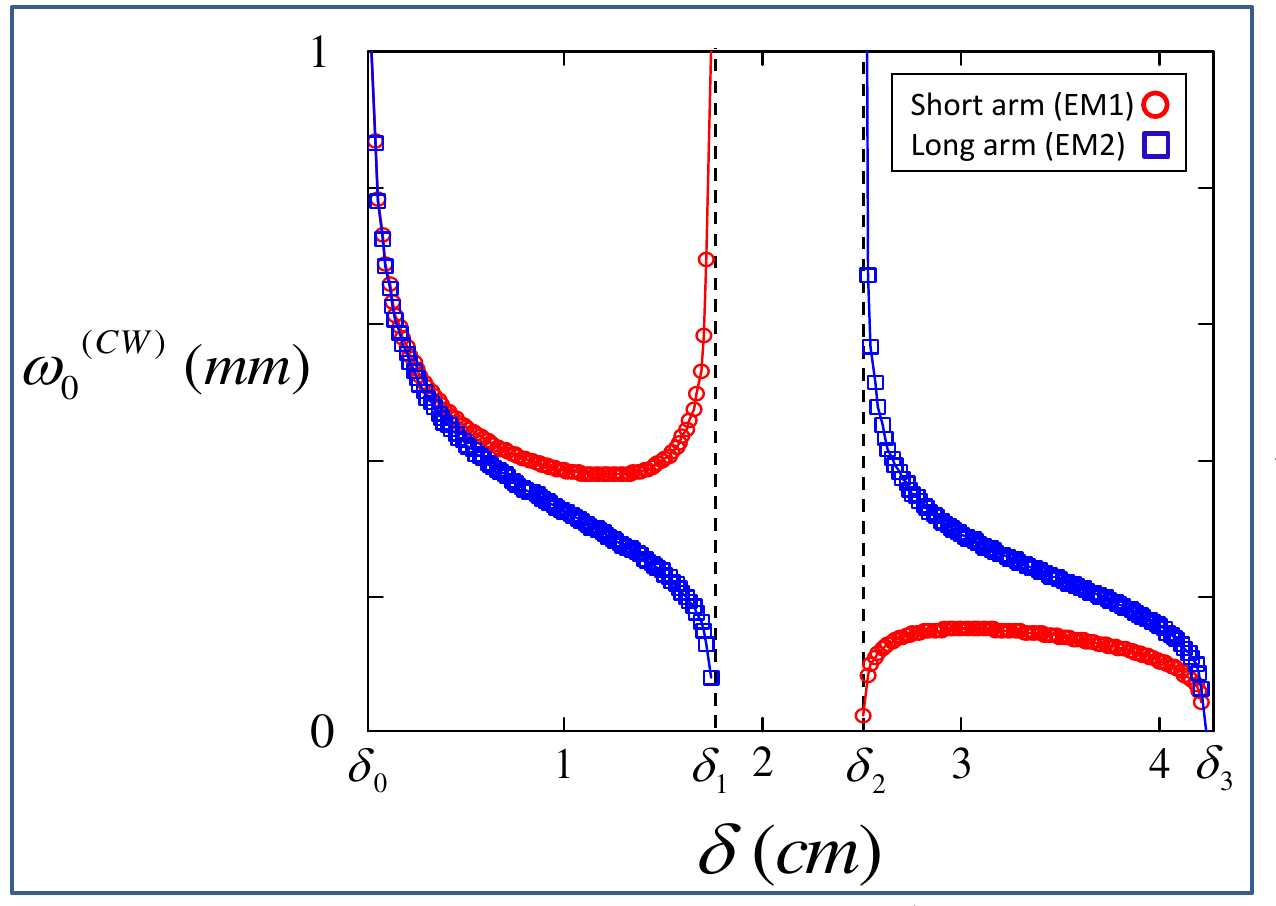}}
 \caption[CW mode size on end mirrors in a non-astigmatic cavity]
 {{Continuous-wave (CW) mode size on the end mirrors, $EM1$ and $EM2$, for the cavity illustrated in Figure~\ref{fig:foldedcavity} with $f_{1}=f_{2}$ = 7.5 cm and arm lengths of $L_{1}$ = 30 cm (short arm) and $L_{2}$ = 40 cm (long arm). At $\delta_{1}$, the beam is collimated in the short arm and focused on $EM2$ in the long arm. This behavior is reversed at $\delta_{2}$.}}
 \label{fig:modeonEM}
\end{figure}

A natural location to calculate the complex beam parameter is at the flat end mirrors, where the mode must arrive with a flat phase front. At the end mirrors, $q_{cw}=iz_{R}$ is purely imaginary, and the mode waist radius, $w_{0}^{(CW)}$, can be plotted as a function of the separation, $\delta$, between the curved mirrors, as presented in Figure~\ref{fig:modeonEM}.

\subsection{Linear Astigmatism} \label{sc.linearastig}

Linear astigmatism is an important consideration in folded cavities with Brewster-cut \mbox{windows/crystals}. Defining the tangential plane as the plane of refraction/reflection (parallel to the main plane of the cavity) and the sagittal plane perpendicular to it, we find that in the tangential plane, the mode size expands (divergence is reduced) as the beam refracts into the Brewster-cut crystal, while in the sagittal plane, the mode size remains the same. This introduces \emph{astigmatism}, since the beam experiences a different geometric path in each plane, given by: $L_{s}=L/n$ for the sagittal plane and $L_{t}=L/n^{3}$ for the tangential plane ($L$ is the thickness of the crystal, and $n$ is the refractive index). This aberration differently shifts the stability limits of the two planes from the non-astigmatic stability limits in Equation~\eqref{eq:stabilitylimits}, resulting in different stability ranges for each plane. Therefore, astigmatism compensation is necessary to ensure stable oscillation in both planes and to obtain a circular Gaussian mode at the end mirror.

The material astigmatism from Brewster-cut windows/crystals can be compensated for using the curved mirrors that fold the beam in the main plane of the cavity. Since the focus of an off axis curved mirror (or lens) is different in the two planes, the folding angles of the curved mirrors introduce an additional astigmatism that (luckily) is of opposite sign to that of a Brewster-cut window, allowing for compensation. The curved mirror focal length for reflection at an angle, $\theta$, is $f_{s}(f,\theta)=f/\cos\theta$ for the sagittal plane and $f_{t}(f,\theta)=f\cos\theta$ for the tangential plane \cite{ABCDfortiltedinterface}, where $f$ is the paraxial focus. Therefore, the mirrors astigmatism $\Delta f(f,\theta)=f_{s}(f,\theta)-f_{t}(f,\theta)$ can compensate for the crystal astigmatism $\Delta L=L_{s}-L_{t}$, by solving:\vspace{-12pt}
\begin{align}
 \label{eq:astig}
 \Delta f(f_{1},\theta_{1})+\Delta f(f_{2},\theta_{2})=\Delta L
\end{align} For $\theta_{1}=\theta_{2}=\theta$ and equal paraxial foci $f_{1}=f_{2}=f$, the solution of Equation~\eqref{eq:astig} is \cite{astigmatismcompensation}:
\begin{align}
 \label{eq:astigsolve}
 \theta=\arccos(\sqrt{1+N^{2}}-N)
\end{align} where $N=(L/4fn)(1-1/n^{2})$.

Note that the solution given in Equation~\eqref{eq:astigsolve} compensates for astigmatism only in the plane-plane configuration, at $\delta=0$. In general, astigmatism can only be compensated for a specific distance, $\delta$, in the cavity. In the Ti:sapphire oscillator, because of the very small pump (and laser) mode size inside the crystal, one usually operates near one of the stability limits, and astigmatism should compensate for that limit. Appendix discusses compensation for the other stability limits.

\section{Cavity Analysis in Mode-Locked Operation} \label{sc.ML.analysis}

In contrast to CW mode analysis, which is straight-forward and fully analytic, the solution of the spatial mode in ML operation is inherently difficult: due to the nonlinearity of the optical Kerr effect, \mbox{the cavity} mode depends on the intensity, which, in turn, depends on the mode, leading to coupling between the two. Therefore, methods to calculate the spatial mode in ML operation are approximated, iterative and mostly non-analytic for the general cavity configuration. In the following, we review several common methods for cavity analysis in ML operation.

\subsection{A Gaussian Beam in a Kerr Medium} \label{sc.beaminkerr}

The additional Kerr lens in the pulsed laser mode (Equation~\eqref{eq:kerrlens}) must be taken into account in order to calculate the Gaussian mode for ML operation. The presence of a lens-like effect inside the crystal, similar to thermal lensing \cite{cavitywithchangelens}, dramatically changes the stability behavior of the cavity. As for the Kerr effect in the laser crystal, several methods have been proposed to solve $q_{ML}$ for ML operation \cite{piche1991,magni1993abcd,lin1999novel}. These methods construct a nonlinear ABCD matrix to treat the propagation of Gaussian beams in the Kerr medium, where the nonlinear matrix depends on the mode intensity inside the crystal. Then, one can solve $q_{ML}$ (for a given peak power of the pulse, $P$) in an iterative manner: starting with the solution of $q_{cw}$ (at $P=0$) as an initial guess, one calculates the nonlinear ABCD matrix based on the mode size in the crystal given by $q_{cw}$; this matrix defines, then, a new resonator with the additional Kerr lens, which yields a new solution of the laser mode, which allows recalculation of the nonlinear matrix, and so forth. The iterations continue, until a steady state is reached for the assumed intra-cavity peak power.

A useful direct approach to solve $q_{ML}$ along this line is to divide the Ti:sapphire crystal into many thin slices, considerably shorter than $z_{R}$, the Rayleigh range of the laser mode. The crystal is then represented as a stack of many thin-lens matrices, where each matrix, $M_{slice}$, consists of a nonlinear lens and propagation through a single slice with thickness $l$:
\begin{align}
 \label{eq:slicematrix}
 \begin{split}
 M_{slice}=
 \begin{bmatrix} 1 & 0 \\ -\frac{1}{f_{NL}(w)} & 1 \end{bmatrix}\begin{bmatrix} 1 & l \\ 0 & 1 \end{bmatrix}=
 \begin{bmatrix} 1 & l \\ -\frac{1}{f_{NL}(w)} & 1 \end{bmatrix}
 \end{split}
\end{align}
where $f_{NL}(w)$ is the nonlinear focus given by Equation~\eqref{eq:kerrlens} for a Kerr medium with thickness $l$ (calculated for the mode size at that slice. One can then calculate $q_{ML}$ in an iterative manner, as described above. Note that the element, $D$, in the matrix, $M_{slice}$, equals $D=1+O(l^{2})$. In the limit, where $l\ll z_{R}$, one can use $D=1$.

Another useful numerical, but not iterative, method was given in \cite{inverseqtransform}, where it was shown that dividing the crystal into many infinitely thin lens-like slices with thickness $dz$ can be represented as a differential equation for the inverse complex beam parameter, $q^{-1}$:
\begin{align}
 \label{eq:kerrdifferential}
 \frac{d}{dz}\left(\frac{1}{q}\right)+\left(\frac{1}{q}\right)^{2}+K {\rm Im}^{2}\left(\frac{1}{q}\right)=0
\end{align}

Here, the nonlinear Kerr lensing process depends on a normalized parameter: $K=P/P_{c}$, where $P$ is the pulse peak power and $P_{c}$ is the critical power for catastrophic self-focusing, given by:
\begin{align}
 \label{eq:criticalpower}
 P_{c}=\frac{\lambda^{2}}{4n_{2}\pi}
\end{align}

\begin{figure}
 \centering
 \centerline{\includegraphics[page=9, clip, trim=0cm 5.1cm 0cm 4.2cm, width=14cm]{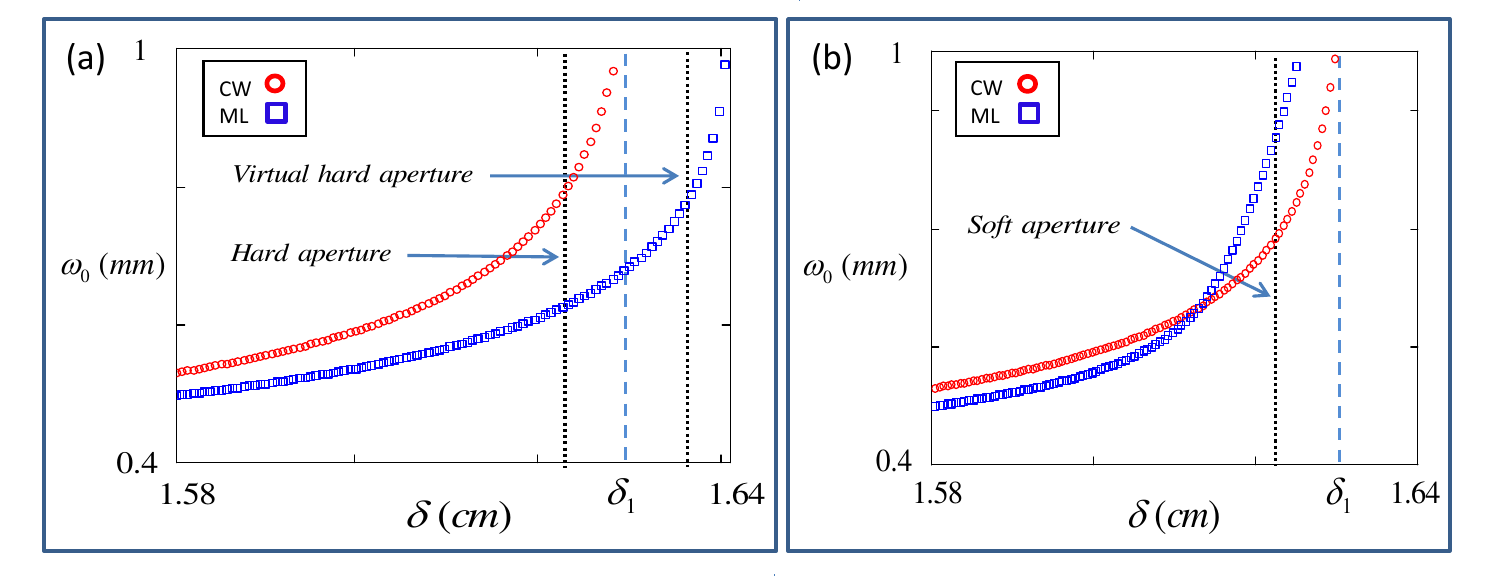}}
 \caption[ML mode size on the short arm end mirror]
 {{Mode-locked (ML) and CW mode size on the end mirror, $EM1$ (short arm), near the CW stability limit, $\delta_{1}$, for the cavity illustrated in Figure~\ref{fig:foldedcavity}. (\textbf{a}) The crystal is at the CW focus with pulse parameter $K\approx0.1$; the ML stability limit is pushed towards higher values of $\delta$ by the Kerr lens. In (\textbf{b}), the crystal is away from focus with pulse parameter $K\approx0.2$; the ML stability limit is pulled down by the Kerr lens to lower values of $\delta$. The dotted-dashed lines (black) represent positions where hard and soft aperture mode-locking techniques can be experimentally employed (these techniques are discussed in detail in Section~\ref{sc:hardaperture}). The long-dashed lines (blue) represent the second stability limit in CW operation.}}
 \label{fig:modeofML}
\end{figure}

Note that the nonlinear process now depends on the normalized parameter $K$. Consequently, other theoretical and experimental values for $P_{c}$ can be defined \cite{criticalpower}. For $K=0$, Equation~\eqref{eq:kerrdifferential} reduces back to propagation of a Gaussian beam in a linear material with refractive index $n$. By scaling the imaginary part of $q^{-1}$ by $\sqrt{1-K}$, the propagation of the new complex beam parameter $\tilde{q}^{-1} = {\rm Re}[q^{-1}]+i {\rm Im}[q^{-1}]\sqrt{1-K}$ through a Kerr medium can be reduced back into a free-space propagation (\emph{i.e.}, K = 0). Thus, the effect of the nonlinearity is fully contained in the transformation of $q^{-1}$, and a problem involving Kerr nonlinearity can be analyzed using linear ABCD matrices, like free space propagation. At the end of the Kerr medium, $\tilde{q}^{-1}$ must be re-transformed by re-scaling the imaginary part of $\tilde{q}^{-1}$ by $1/\sqrt{1-K}$. Since there is no ABCD matrix that represents this transformation, one can no longer represent the entire cavity by a single ABCD matrix. $q_{ML}$ of the stable mode must, thus, be obtained by a numerical solution of the stability condition $|A+D|<2$ for the general cavity configuration illustrated in Figure~\ref{fig:foldedcavity}. Analytical solutions, however, can still be achieved for a ring cavity \cite{stabilitydesign} or for a symmetrical linear cavity \cite{analyticlindesign} (equal cavity arms). The result of the calculation is a different mode size for $w_{0}^{(ML)}$ as a function of $\delta$ compared to $w_{0}^{(CW)}$ (see Figure~\ref{fig:modeofML}).

It is important to note that all calculations are performed within the aberration-free approximation for the Kerr lens, in which the transverse variation of the refractive index is approximated to be parabolic, so that the beam maintains its Gaussian shape during propagation, and the ABCD method to analyze the cavity can be applied. A detailed discussion regarding the limits of the aberration-free approximation can be found in \cite{abberationapp}. Note, in addition, that in Figure~\ref{fig:modeofML}, astigmatism was neglected for simplicity. \mbox{The effects of} nonlinear astigmatism on the ML mode are discussed in Section~\ref{sc.additionastig}.

The solution for the ML mode size presented in Figure~\ref{fig:modeofML} highly depends on the Kerr medium (crystal) position. Changing the position of the crystal along the beam affects the mode intensity in the crystal and, hence, the nonlinear response, leading to a different (sometimes dramatically) solution of $w_{0}^{(ML)}$. In Figure~\ref{fig:modeofML}a, the crystal was positioned at the CW focus point between the curved mirrors, whereas in Figure~\ref{fig:modeofML}b, the crystal is located away from the focus, showing the completely opposite behavior. While in Figure~\ref{fig:modeofML}a, the Kerr lens pushes the stability limit for ML operation towards higher values of $\delta$, \mbox{in Figure~\ref{fig:modeofML}b}, the stability limit is pulled towards lower values. The position of the crystal accordingly determines which method will be employed in order to mode-lock the cavity (hard or soft aperture), \mbox{as will be discussed} in Section~\ref{sc:hardaperture}.

It is important to note that the above method (\cite{inverseqtransform} and Equation~\eqref{eq:kerrdifferential}) assume a circular mode inside the crystal and do not include the effects of the more realistic elliptical (astigmatic) mode in the Brewster-cut crystal. One may try to calculate $q_{ML}$ for each plane separately by assuming a circular mode with the corresponding mode size ($w_{s}/w_{t}$ for the sagittal/tangential plane, respectively), but this is oversimplified in most cases. While such a simple separation of the planes may produce a qualitative understanding of the mode, it cannot provide a quantitative solution, since the planes in reality are coupled by the Kerr lensing effect (the mode size in one plane affects the peak intensity, which then affects the mode of the other plane). A detailed description of the coupled propagation equations of a Gaussian beam through a Kerr medium is provided in \cite{astigmaticselffocusing}. We further discuss the effects of nonlinear astigmatism in Sections~\ref{sc.additionastig} and \ref{Sc.nononlinearastig}.

\section{Mode-Locking Techniques} \label{sc:hardaperture}

The addition of the Kerr lens changes the laser mode for pulses, but does not necessarily impose a preference for pulsed operation. This preference is achieved when a proper aperture in the cavity selectively prefers the pulsed mode over the CW mode. Here, two methods are common to induce losses on the CW mode compared to the pulsed mode: hard aperture and soft aperture. In the following, we provide an intuitive and quantitative description of these methods.

\subsection{Hard/Soft Aperture Mode-Locking}

In order to prefer ML operation, one can exploit the variation of $w_{0}^{(ML)}$ compared to $w_{0}^{(CW)}$ (Figure~\ref{fig:modeofML}) in combination with an aperture to introduce an intensity-dependent loss mechanism into the cavity. Usually, the working point for mode-locking is near a stability limit, where the laser mode is tightly focused into the crystal, maximizing the intensity-dependent Kerr effect. At a given $\delta$ close to a stability limit (Equation~\eqref{eq:stabilitylimits}), the crystal position can be translated along the beam propagation axis to affect the Kerr strength, defined by:
\begin{align}
\gamma=\frac{P_{c}}{w_{cw}}\left(\frac{dw}{dP}\right)_{P=0} \label{eq:gamma}
\end{align} where $P$ is the pulse peak power and $w$ is the mode radius at one of the end mirrors. The Kerr strength of Equation~\eqref{eq:gamma} represents the change of the mode size due to a small increase in $P$ (normalized to the CW mode size), which is a measure for the effect of the Kerr lens on the cavity. Investigating the variation of $\gamma$ as a function of the crystal position, one finds two possible mechanisms to inflict loss on the CW mode \cite{KLM}: hard aperture, which is applicable when the ML mode is smaller than the CW mode, as in Figure~\ref{fig:modeofML}a; and soft aperture, which is applicable in the opposite scenario when the ML mode is larger than the CW mode, as in Figure~\ref{fig:modeofML}b.

\subsubsection{Hard Aperture}

\begin{figure}
 \centering
 \centerline{\includegraphics[page=1, clip, trim=0cm 5.1cm 0cm 3.4cm, width=14cm]{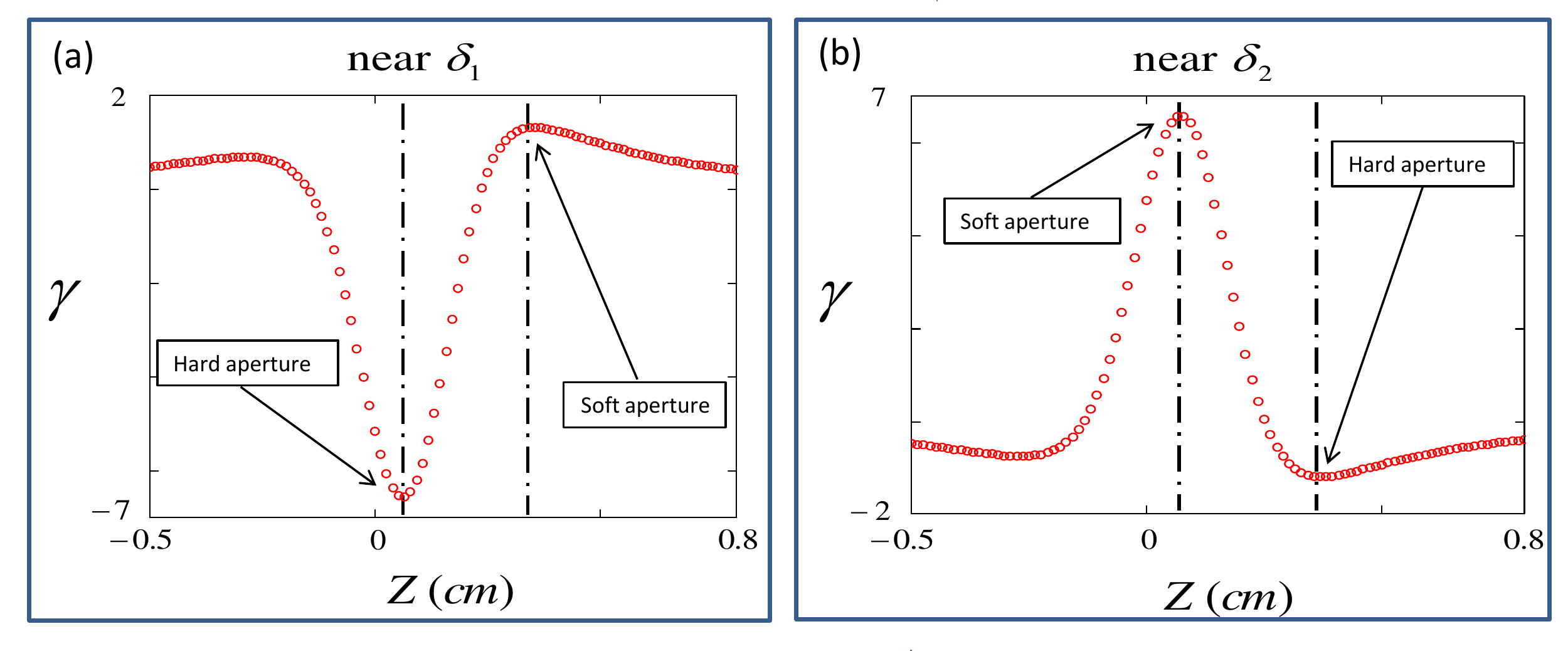}}
 \caption[Kerr strength vs. crystal position]
 {{(\textbf{a}) Kerr strength $\gamma$ at $EM1$ as a function of the crystal position, $Z$, near the stability limit, $\delta_{1}$. At $Z=0$, the crystal center is located at a distance $f_{1}$ from $M1$. Positive values of $Z$ correspond to the translation of the crystal away from mirror $M1$. (\textbf{b}) Kerr strength $\gamma$ as a function of the crystal position, $Z$, at $EM2$ near the stability limit, $\delta_{2}$. At $Z=0$, the crystal center is located at a distance, $f_{2}$, from $M2$. Positive values of $Z$ correspond to the translation of the crystal away from $M2$.}}
 \label{fig:gammavsZ}
\end{figure}

Figure \ref{fig:gammavsZ}a plots $\gamma$ (calculated at the end mirror, $EM1$) as a function of the crystal position near the end of the first stability zone, $\delta\lesssim\delta_{1}$. When the crystal is located near the focus, $\gamma$ is negative, and hence, the Kerr lensing reduces the mode size for ML compared to CW operation. This enables the introduction of a physical (``hard'') aperture near the end mirror, $EM1$, to employ loss on the CW mode, while the smaller ML mode passes through the aperture without attenuation. Hard aperture is illustrated in Figure~\ref{fig:modeofML}a, where the power-dependent stability limit, $\delta_{1}$, for ML operation is pushed towards higher values of $\delta$. Hence, the working point for mode-locking is at $\delta\lesssim\delta_{1}$, where the CW mode is larger than the ML mode.

\subsubsection{Soft Aperture}

As the crystal is translated away from the focus towards $M2$, $\gamma$ becomes positive, thereby providing a mode size for ML operation, which is larger than the CW mode. In this situation, the stability limit for ML operation is pulled down to lower values of $\delta$ by the added Kerr lens. Therefore, a physical aperture cannot be used at $EM1$. However, the larger mode size for ML at $EM1$ corresponds to a smaller mode size at the crystal compared to the CW mode. This enables the pump beam inside the crystal to be used as a ``soft'' aperture, by setting a small pump mode to overlap with the smaller ML mode at the focus. Consequently, the pump mode acts as a ``soft aperture'', in which the CW mode suffers from poor pump-mode overlap compared to the ML mode. As can be seen from Figure~\ref{fig:gammavsZ}a, the soft aperture technique for mode-locking near $\delta_{1}$ is less efficient, because the translation of the crystal away from the focus increases the mode size in the crystal, lowering the value of $|\gamma|$ and also increasing the laser threshold. The efficiency of soft aperture compared to hard aperture mode-locking is better when mode-locking near the second stability zone, $\delta\gtrsim\delta_{2}$, as shown in Figure~\ref{fig:gammavsZ}b, where $\gamma$ is plotted as a function of the crystal position on the end mirror of the long arm ($EM2$). It is therefore beneficial to mode-lock with hard aperture near $\delta_{1}$ and with soft aperture near $\delta_{2}$.

\subsection{Virtual Hard Aperture Mode-Locking} \label{ch:passivehardaperture}

\begin{figure}
 \centering
 \centerline{\includegraphics[page=2, clip, trim=0cm 0cm 0cm 0.2cm, width=13cm]{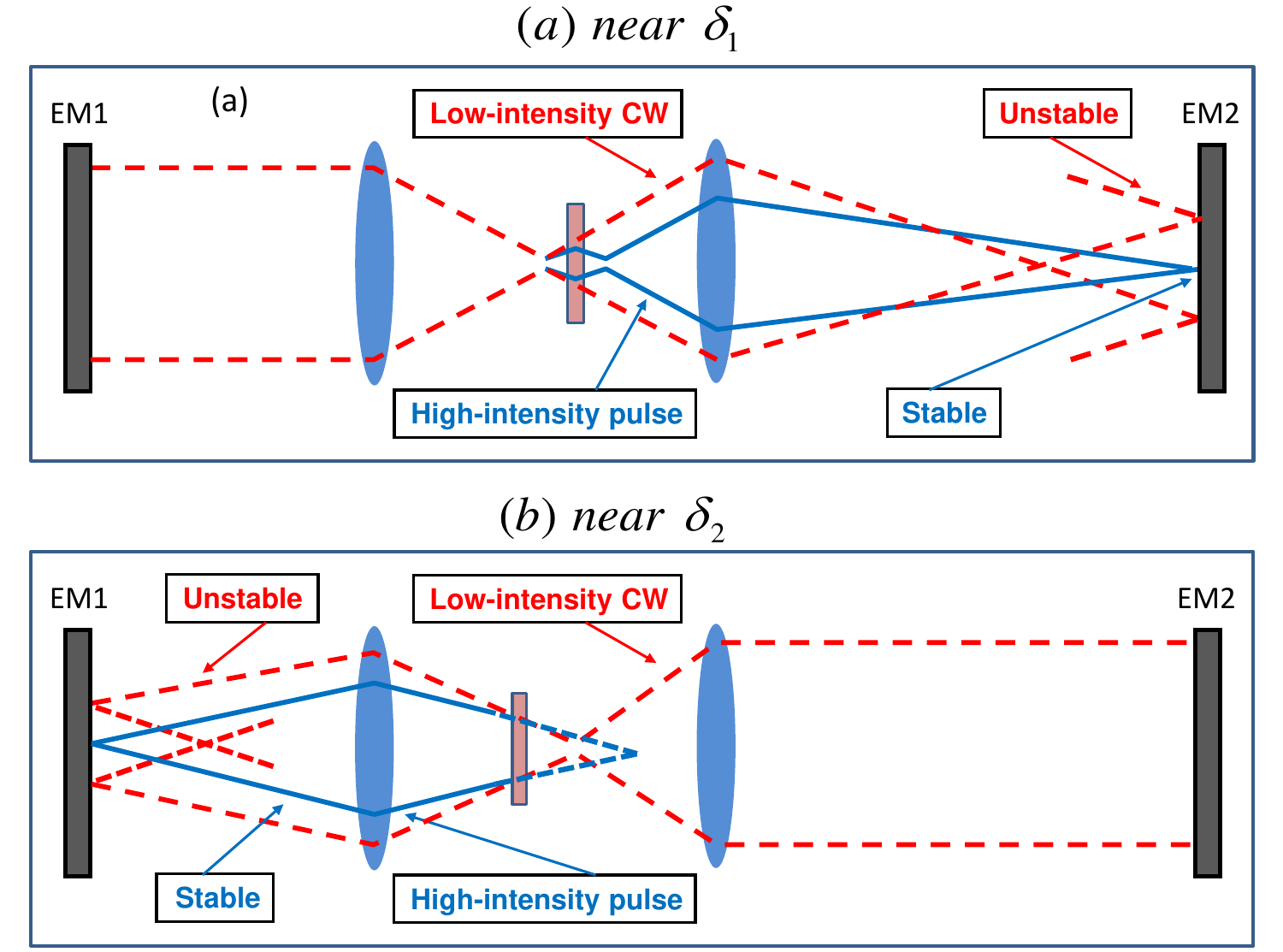}}
 \caption[Kerr lensing as a mode-locking stabilizing mechanism]
 {{Illustration of the Kerr lensing effect as a mode-locking stabilizing mechanism in virtual hard aperture technique: (\textbf{a}) increasing $\delta$ beyond the stability limit, $\delta_{1}$, destabilizes CW operation, while ML operation remains stable as the effective distance between $f_{1}$ and $f_{2}$ becomes shorter, by the nonlinear forward imaging of the focus point; (\textbf{b}) decreasing $\delta$ below the stability limit, $\delta_{2}$, destabilizes CW operation, while ML operation remains stable as the effective distance between $f_{1}$ and $f_{2}$ becomes larger by nonlinear virtual imaging of the focus point.}}
 \label{fig:kerrlensing}
\end{figure}

Maybe the most convenient technique for hard aperture mode-locking near the first stability limit, $\delta_{1}$, without using a physical aperture, is to choose the mode-locking point at $\delta\gtrsim\delta_{1}$, a little bit beyond the stability limit, as was experimentally demonstrated in \cite{MLnoCW}. As can be seen in Figure~\ref{fig:modeofML}a, crossing the stability limit, $\delta_{1}$, affects only CW operation, while the cavity can still be stable for ML operation. At $\delta\gtrsim\delta_{1}$, though the cavity is unstable for CW, it will operate very well, only with an elevated threshold, due to increased diffraction losses. Diffraction losses are increased as $\delta$ is increased beyond $\delta_{1}$. The concept of re-stabilization is illustrated in Figure~\ref{fig:kerrlensing}a. The Kerr lens images the focus of the ML mode forward, effectively shortening the distance between $M1$ and $M2$ compared to CW, thereby pushing the stability limit for mode-locking towards higher values of $\delta$. The unstable CW passively suffers from diffraction losses, which is effectively equivalent to employing an aperture. Mode-locking can still be initiated by knocking on one of the end mirrors, just as with standard hard/soft aperture. On a practical note, thermal lensing in the Ti:sapphire crystal must be considered, as it may also shift the stability limits of the cavity. In the experiments shown in \cite{MLnoCW}, the absence of thermal lensing was experimentally verified. Virtual hard aperture mode-locking can be achieved also near $\delta_{2}$ (though far less effectively than soft aperture) by choosing the mode-locking point at $\delta\lesssim\delta_{2}$. As illustrated in Figure~\ref{fig:kerrlensing}b, CW operation is destabilized, while for ML operation, the cavity can still be stable. The Kerr lens then virtually images the focus of the ML mode backwards, thereby effectively lengthening the distance between $M1$ and $M2$ for the ML mode compared to CW. Thus, one can choose a mode-locking point at $\delta\lesssim\delta_{2}$, which is stable for ML operation, but not stable for CW operation. This, however, is much less efficient than mode-locking at $\delta\gtrsim\delta_{1}$, since it requires a weaker Kerr lensing effect in order to create a virtual image of the focus point rather than a real image.

Last, we note that hard/soft aperture mode-locking can be achieved in the same manner as explained above near the stability limits, $\delta_{3}$ and $\delta_{0}$. At $\delta_{3}$, hard aperture has the stronger nonlinear response, similar to $\delta_{1}$. However, due to experimental considerations, one usually prefers to have a collimated beam at one of the cavity arms, which is not the case at $\delta_{3}$ (see Figure~\ref{fig:geometricMODES}). At $\delta_{0}$, soft aperture has the stronger nonlinear response, similar to $\delta_{2}$, yet cavity operation near the plane-plane limit, $\delta_{0}$, is usually less convenient compared to the point-plane limit, $\delta_{2}$, since it is more sensitive to the misalignment of the end mirrors.

\section{Nonlinear Kerr Lens Astigmatism} \label{sc.additionastig}

\begin{figure}
 \centering
 \centerline{\includegraphics[page=10, clip, trim=0cm 4.6cm 0cm 3.9cm, width=15cm]{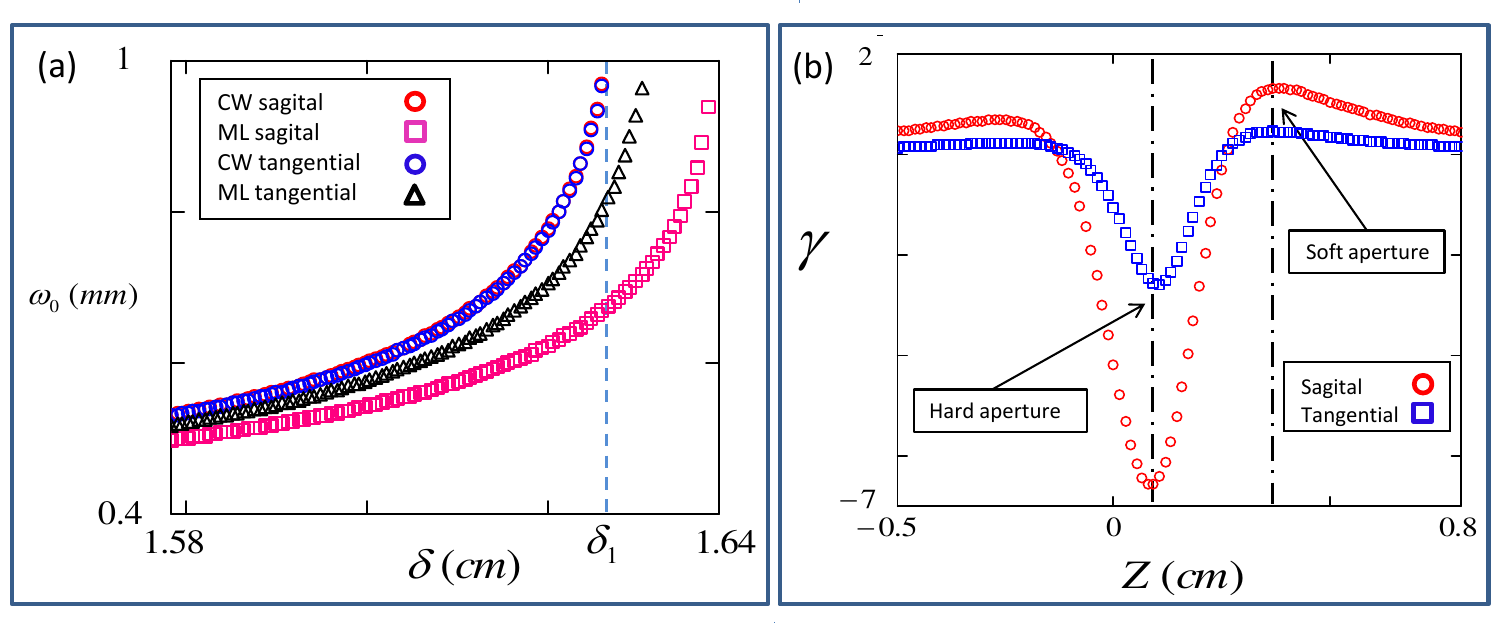}}
 \caption[Nonlinear astigmatism]
 {{(\textbf{a}) ML and CW mode size for both sagittal and tangential planes on the end mirror, $EM1$ (short arm), near the CW stability limit, $\delta_{1}$, for hard aperture mode-locking; \mbox{(\textbf{b}) Kerr strength} $\gamma$ at $EM1$ as a function of the crystal position, $Z$, near the stability limit, $\delta_{1}$.}}
 \label{fig:kerrastigplots}
\end{figure}

\begin{figure}
 \centering
 \centerline{\includegraphics[page=11, clip, trim=2.1cm 1.6cm 2.1cm 1.4cm, width=9cm]{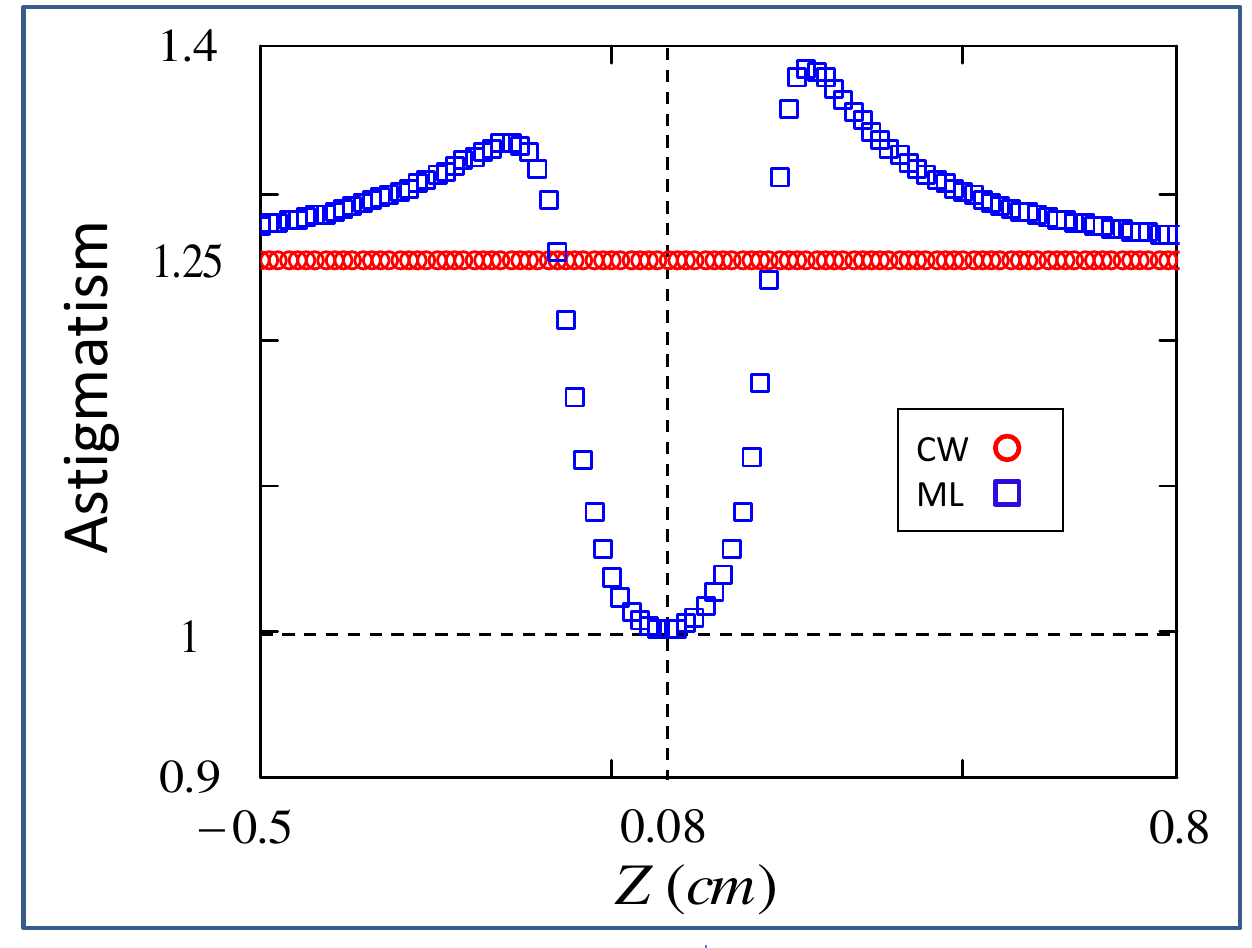}}
 \caption[Power dependent nonlinear astigmatism compensation]
 {{CW and ML astigmatism (defined as $w_{s}/w_{t}$) on the end mirror, $EM1$, as a function of the crystal position for hard aperture mode-locking at $\delta\lesssim\delta_{1}$. A linear astigmatism of $\approx$$1.24$ is pre-introduced into the CW mode, allowing for the stronger Kerr effect in the sagittal plane to ``catch up'' with the weaker tangential plane, resulting in the astigmatically-compensated ML mode at $Z\approx0.08$ for a specific value of $K=0.1$.}}
 \label{fig:kerrastigcomp}
\vspace{-12pt}
\end{figure}

Kerr lensing in the Brewster-cut Ti:sapphire crystal introduces an additional source of nonlinear astigmatism into the cavity. As the beam refracts into the crystal, the mode size in the tangential plane, $w_{t}$, expands, while the sagittal mode size, $w_{s}$, remains the same, hence reducing the intensity and the nonlinear response in the tangential plane. The difference in the nonlinear response between the two planes will produce an intensity-dependent astigmatism, even if the linear astigmatism is fully corrected. In terms of the power-dependent stability limit, $\delta_{1}$, of mode-locked operation illustrated in Figure~\ref{fig:kerrastigplots}a, both the sagittal and tangential stability limits will be pushed towards higher values of $\delta$, but the sagittal limit will be pushed farther away, compared to the tangential limit. Note that only hard aperture mode-locking pushes the stability limits towards higher values of $\delta$, in contrast to soft aperture mode-locking, where the stability limits are pushed towards lower values of $\delta$. However, in both hard and soft aperture techniques, the relative variation of the power-dependent stability limit will be higher in the sagittal plane compared to the tangential plane. In terms of the Kerr lens strength, $\gamma$, plotted in Figure~\ref{fig:kerrastigplots}b, both the tangential and sagittal planes will have a similar qualitative behavior, but the absolute values of $\gamma$ will be reduced in the tangential plane compared to the sagittal. Therefore, a fully-corrected CW mode will not remain corrected after mode-locking. The standard solution to the problem is to pre-compensate for the nonlinear astigmatism \cite{stabilitydesign} by introducing extra linear astigmatism for the CW mode in the opposite direction, as seen in Figure~\ref{fig:kerrastigcomp}, such that the plane with the stronger $|\gamma|$ will ``catch up'' with the weaker one. By changing the values of the angles, $\theta_{1}$ and $\theta_{2}$, away from perfect linear astigmatic correction, one can pre-compensate for nonlinear astigmatism at $\delta_{1}$. Consequently, a deliberately non-circular CW mode will become circular after mode-locking. Note, however, that this compensation holds only for a specific value of $K=P/P_{c}$, \emph{i.e.}, specific intra-cavity peak power. Increasing (lowering) $K$ with the same folding angles (\emph{i.e.}, the same linear astigmatism) will result in over (under) compensating for the nonlinear astigmatism. Any change in parameters that keeps the CW astigmatism compensated for, but that affects the intra-cavity intensity, be it the peak power or mode size in the crystal, will require a change in the folding angles to match the precise CW astigmatism needed to converge into a non-astigmatic ML beam. This includes a change in: pump power, pump focusing, output coupler, short arm length and, also, $Z$ or $\delta$. This requires specific compensation for every time one changes cavity parameters, making nonlinear astigmatism a nuisance in standard cavity designs. Recently, a novel cavity design has been demonstrated that nulls nonlinear astigmatism completely \cite{nononlinastig}, as will be elaborated in Section~\ref{Sc.nononlinearastig}.

\section{Dispersion Compensation} \label{sc.dispersion}

In order for an optical cavity to sustain a pulse, the temporal shape and duration of the pulse must remain stable as it circulates through the cavity. Since the Ti:sapphire crystal and other optical elements in the cavity are dispersive, the pulse is deformed as it passes through them, due to the wavelength dependence of the refractive index. Thus, dispersion compensation is imperative in order to sustain short pulses. In the following, we provide a concise survey of the basic principles of dispersion management and the most commonly used dispersion compensation devices of prism pairs and chirped mirrors. For a more detailed study of the topic, readers are referred to the literature \cite{dispersionreview,Diels@shortpulse,disperdevice}.

\subsection{Group Delay Dispersion}

When a pulse propagates through a dispersive material, its spectral phase, $\phi(\omega)$, is affected. By expanding $\phi(\omega)$ in a Taylor series around the pulse central frequency, $\omega_{0}$, one can identify three major effects, corresponding to the first three terms in the series: (1) overall phase accumulation; the propagation phase added to all frequencies; (2) group delay (GD); the entire pulse is delayed compared to a pulse propagating in free space; and (3) group delay dispersion (GDD); a frequency-dependent group delay of the different spectral components of the pulse. Since GDD (not GD) causes a temporal broadening of the pulse in every round trip through the cavity \cite{pulsebroadening}, it must be compensated for to sustain the pulse in the cavity over time. In terms of the wavelength, $\lambda$, the GDD of a medium is given by:
\begin{align}
 \label{eq:GDDmaterial}
 {\rm GDD}=\frac{d^{2}\phi}{d\omega^{2}}=\frac{\lambda^{3}}{2\pi c^{2}}\frac{d^{2}}{d\lambda^{2}}\rm OP(\lambda)
\end{align}
where $\rm OP(\lambda)=n(\lambda) L(\lambda)$ is the optical path in the cavity of wavelength $\lambda$. Note that pulse broadening can also be measured in terms of the group velocity dispersion (GVD) which is the GDD per millimeter of the corresponding material. To calculate the GDD, due to a window with thickness $L$ inside the cavity, we can assume $L$ to be constant for all wavelengths (this is not exact for Brewster windows, but a very good approximation). For the Ti:sapphire crystal in the NIR range of the spectrum, Equation~\eqref{eq:GDDmaterial} results in a positive GDD for material components. For simplicity, higher order terms in the phase expansion, which correspond to third-, fourth- and fifth-order dispersion effects, are neglected so far; yet, the higher order terms (along with the overall gain bandwidth) are eventually the limiting factor for the shortening mechanism of the pulse duration. Expressions for higher order dispersion terms can be found in \cite{FODandFOD}.

\subsection{Chirped Mirrors}

\begin{figure}
 \centering
 \centerline{\includegraphics[page=4, clip, trim=0cm 0cm 0cm 0cm, width=10.5cm]{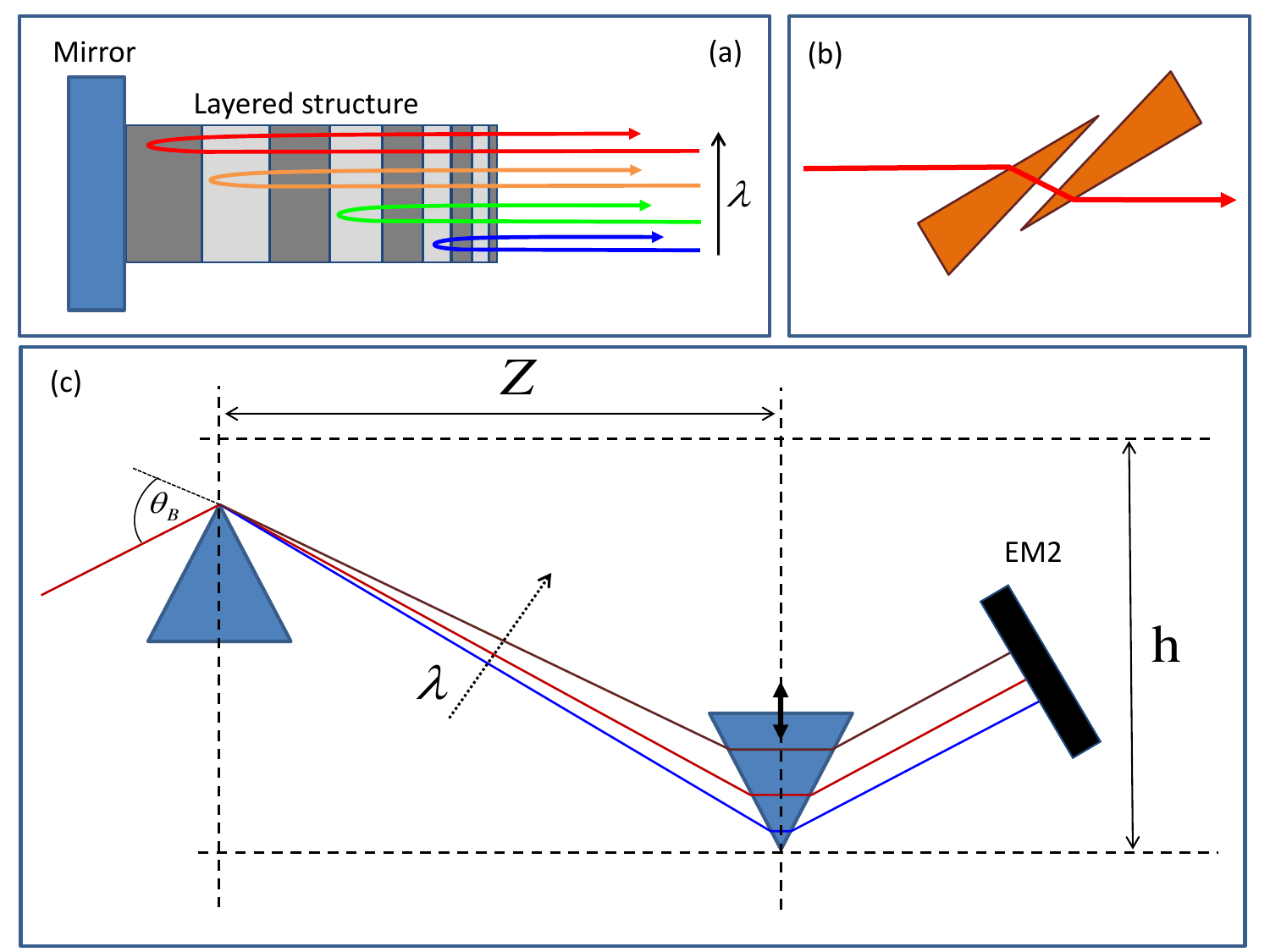}}
 \caption[Dispersion compensation devices]
 {{Dispersion compensation devices: (\textbf{a}) chirped mirrors; (\textbf{b}) a pair of wedged windows; \mbox{(\textbf{c}) a pair of prisms}.}}
 \label{fig:dispersdevices}
\end{figure}

In order to maintain near zero net GDD in the cavity, components of tuned negative GDD must be incorporated in the cavity. Chirped mirrors are a common component for dispersion \mbox{compensation \cite{chirpedmirrors,chirpedmirrorreview}}, which are coated with a stack of dielectric layers, designed such that different wavelengths penetrate a different depth in the stack, as illustrated in Figure~\ref{fig:dispersdevices}a. Mirrors are specified in the amount of negative GDD per bounce they provide. In many cases, a single mirror has a strong GDD oscillation across the spectrum, and pairs of mirrors are commonly designed with opposite oscillations, such that the combined GDD of the pair is spectrally smooth. The technology of dielectric coatings for the manipulation of ultrashort pulses has matured in recent years, and now, even double-chirped mirrors are available for Ti:sapphire cavities that provide specifically-tuned negative GDD to also compensate for higher order dispersion over an extremely wide spectral range, allowing one to achieve pulses with an extreme bandwidth (octave-spanning spectrum) \cite{DCM1,DCM2,DCM3}. Since compensation with chirped mirrors is discrete in nature (a finite dispersion per bounce on the mirror), one can use a wedge window pair to fine-tune the dispersion, as illustrated in Figure~\ref{fig:dispersdevices}b. By controlling the insertion of one window, the variable thickness adds positive GDD in a controlled manner, allowing for continuous compensation.

\newpage

\subsection{Prism Pair} \label{sc.prismpair}

In contrast to material dispersion (where only the refractive index, $n$, is wavelength-dependent), one can also introduce geometric dispersion in which the geometrical path, $L(\lambda)$ (Equation~\eqref{eq:GDDmaterial}), is also wavelength-dependent. Geometric dispersion can be introduced using a prism pair, illustrated in Figure~\ref{fig:dispersdevices}c, in either one of the cavity arms. The general concept of a prism pair is to manipulate the optical path of different frequencies in such a way that all the frequencies will experience the same cavity round-trip time. The relative time delay between the frequencies caused by the prism pair compensates for the time delay caused by other dispersive material in the cavity (e.g., the Ti:sapphire crystal). \mbox{The result is} that a prism pair can generate both negative and positive GDD in a controlled, tuned manner \cite{prismpairapprox,fork@prismpair}. The GDD that the pulse experiences as it passes through the prism pair can be calculated using Equation~\eqref{eq:GDDmaterial}, and the optical path, $\rm OP(\lambda)$, in the entire prism pair system (air and material) is calculated using geometrical optics \cite{GDDprismpairexpression,prismpairEXACT,arbitraryprismpair}, resulting in:
\begin{align}
 \label{eq:GDDprismpair}
 {\rm GDD}=-\frac{2\lambda}{\pi c^{2}}\left(\lambda\frac{dn}{d\lambda}\right)^{2}R_{p}+\left(\frac{\lambda^{3}}{\pi c^{2}n^{2}}\frac{d^{2}n}{d\lambda^{2}}\right)H
\end{align}
where $R_{p}$ is the separation between the prisms tips, $H$ is the second prism penetration and $n$ is the refractive index of the prisms. Therefore, the net GDD that the pulse experiences as it passes through a prism pair can be continuously controlled by changing $R_{p}$ or $H$. Note that increasing the separation between the prisms always introduces negative GDD, while increasing the prism penetration can produce both positive and negative GDD. For the Ti:sapphire crystal in the NIR spectrum, increasing the prism penetration corresponds to positive GDD. Optimization guidelines for a prism pair setup can be found in \cite{dispersionmap}, using a dispersion map in which the second and third order dispersions are plotted as orthogonal coordinates. The resulting dispersion vector can be compensated for by optimizing the prism pair parameters (separation, penetration and prism material) to reduce the dispersion vector to zero. Similarly, a grating pair can be used for larger values of controlled negative dispersion \cite{gratingpair} at the expense of increasing losses. The combination of a prism and a grating in a single device (termed: ``grism") was demonstrated for higher order dispersion management \cite{grism}. One can also use a single prism and a wedged mirror to compensate for dispersion \cite{prismandwedge}.

In this regard, one should be aware of the additional linear astigmatism introduced by prisms (or wedge windows): because a prism pair is effectively a Brewster window separated in two parts, \mbox{it produces} astigmatism (see Section~\ref{sc.linearastig}), which is generally much smaller than that of the crystal. This added astigmatism was not included while solving Equations~\eqref{eq:astig} and \eqref{eq:genastig} and will require fine-tuning of the folding angles of the curved mirrors to re-compensate for the overall astigmatism in the cavity. Note, however, that astigmatism from a prism pair exists only if the beam passing through the pair is not collimated (\emph{i.e.}, the pair is placed in the long arm while mode-locking near $\delta_{1}$ or in the short arm while mode-locking near $\delta_{2}$). This is because, for a collimated beam, the pair astigmatism, $\Delta L$, is completely negligible compared to the long Rayleigh range of the beam.

\section{Typical Characteristics of Pulsed Operation} \label{sc.Ex.analysis}

\begin{figure}
 \centering
 \centerline{\includegraphics[page=13, clip, trim=0cm 0cm 0cm 0cm, width=14cm]{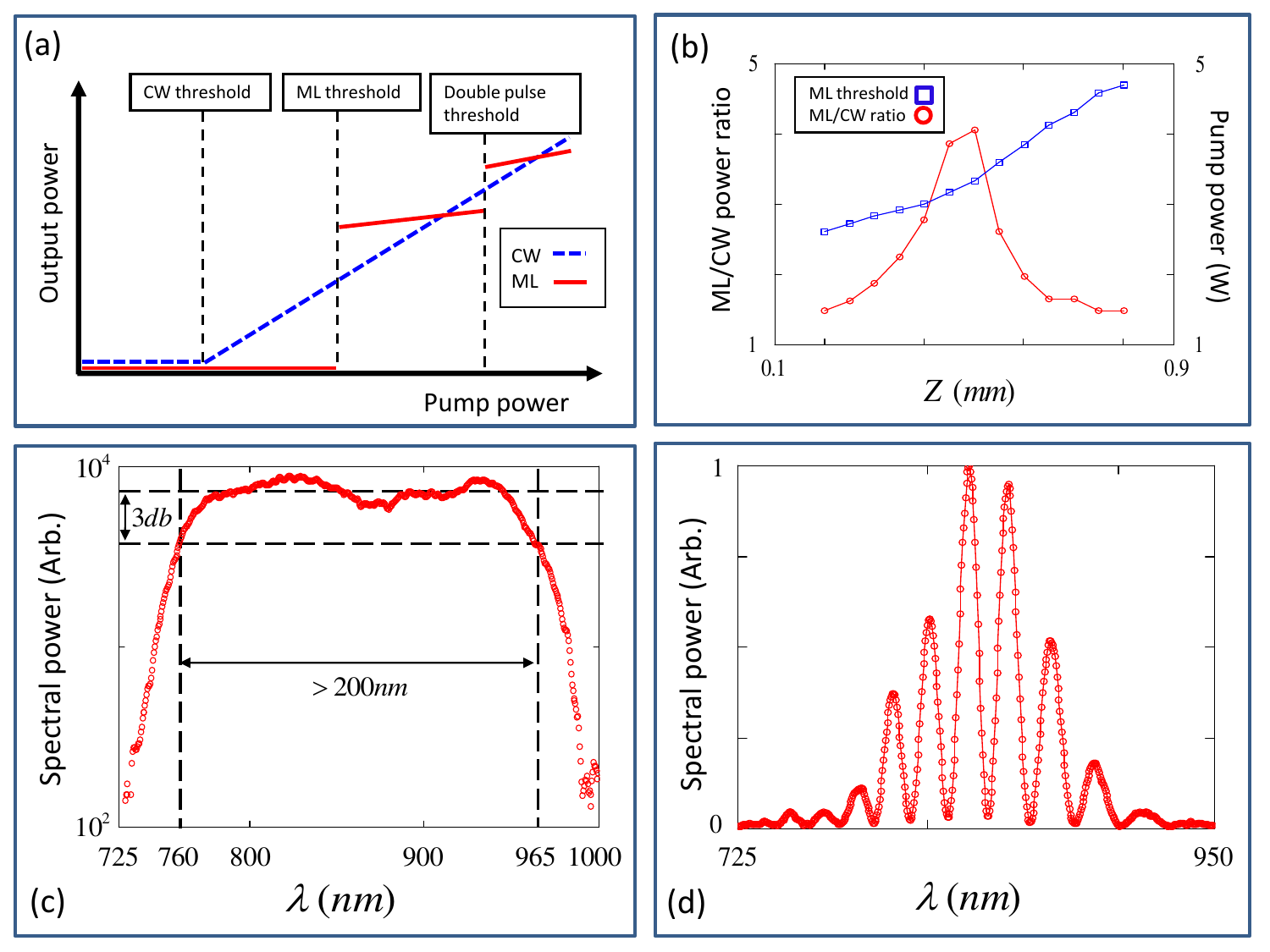}}
 \caption[Qualitative mode-locking behavior]
 {{(\textbf{a}) Qualitative behavior of pulsed operation for a given mode-locking point at $\delta\gtrsim\delta_{1}$. (\textbf{b}) ML threshold and ML-to-CW ratio of output powers as a function of $Z=\delta-\delta_{1}$, while mode-locking with the virtual hard aperture technique at $\delta>\delta_{1}$. Measurements were taken from a typical configuration of a Ti:sapphire cavity, including a $3$ mm-long Brewster-cut crystal, curved mirrors of $f_{1}=f_{2}$ = 5 cm, $L_{1}$ = 20 cm and $L_{2}$ = 75 cm, an output coupler of $95\%$ reflectivity and the pump focused to a diameter of $\approx$$22$ $\upmu$m in the Ti:sapphire crystal. \mbox{(\textbf{c}) Typical} measured spectrum (not optimized) of an ultrashort pulse with a bandwidth of $>$200 nm at full width at half maximum (FWHM). (\textbf{d}) Measured spectrum in the double pulse regime where the spectral interference pattern can be~observed.}}
 \label{fig:MLbehavior}
\end{figure}

The spectroscopic and laser characteristics in CW operation were well studied in the past and can be found in \cite{TiSmedium}. Pulsed operation of the Ti:sapphire laser was also extensively studied, leading to remarkable results in terms of pulse duration \cite{shortestpulse}, repetition rate \cite{highreprate}, average power \cite{highoutputpower}, mode-locking threshold \cite{ultralowthreshold} and pulse energy \cite{higheenergypulse}. Here, we provide some of the basic characteristics and typical qualitative behavior of a Ti:sapphire oscillator in ML operation, especially focused on the onset of ML and the experimental procedure to obtain ML. Using virtual hard aperture mode-locking (Section~\ref{ch:passivehardaperture}), pulsed operation can be achieved in a band of $\delta$ values slightly beyond the stability limit, $\delta_{1}$, showing a qualitative behavior, as plotted in Figure~\ref{fig:MLbehavior}a. Pulsed operation can be achieved only after the pump power is raised beyond the CW threshold up to a certain ML threshold value, where the appearance of pulses is abrupt in terms of pump power. The threshold-like behavior of mode-locking was elegantly explained, both theoretically and experimentally, as a first order phase transition by the theory of statistical light mode dynamics \cite{SLD@theory2,SLD@exp2}.


At the ML threshold, the output power of the pulse will be higher than the CW output power, since the pulse enjoys lower diffraction losses than the CW. Figure \ref{fig:MLbehavior}b plots the ratio between ML output power and the CW output power as $\delta$ is increased beyond the stability limit ($\delta\gtrsim\delta_{1}$). The ratio $\gamma_{e}=P_{ML}/P_{CW}$ is an experimental measure of the Kerr lensing strength, similar to $\gamma$ of Equation~\ref{eq:gamma}. In addition, the mode-locking threshold as a function of $\delta$ is also plotted in Figure~\ref{fig:MLbehavior}b. One finds that both the ML threshold and the ML power increase monotonically with increasing $\delta$, due to the need to overcome increasing diffraction losses, but the Kerr lensing strength has a maximum efficiency point, where ML is optimal.

Figure \ref{fig:MLbehavior}c plots a typical measured broadband spectrum of an ultrashort pulse at the point of maximum Kerr lensing efficiency. The typical bandwidth of $>200 nm$ at FWHM can be achieved even without the use of double-chirped mirrors or additional bandwidth maximization techniques (these techniques will be discussed in Section~\ref{sc.optimization}). Several techniques for temporal characterization of the pulse were developed in the past, such as interferometric autocorrelation (IAC), frequency-resolved optical gating (FROG) and spectral phase interferometry for direct electric-field reconstruction (SPIDER) \cite{IAC,FROG,SPIDER,SHGFROG}. For a given mode-locking point at $\delta\gtrsim\delta_{1}$, further increasing of the pump above the ML threshold, as illustrated in Figure~\ref{fig:MLbehavior}a, will increase the CW output power, while the ML output power will remain approximately the same. A qualitative and intuitive explanation for this phenomenon is that the soliton-like pulse peak power is ``quantized'', \emph{i.e.}, a unique peak power stabilizes the cavity for ML operation, to which the laser clutches. As the pump power is further increased, a small increase in ML power can be observed, but the CW power catches up quickly. When the CW output power becomes larger than the pulse output power, the excess energy excites a CW mode, which oscillates in the cavity along with the pulse, generating an output spectrum of a broad bandwidth with a narrow CW spike attached to it. The peak power ``quantization'' effect is manifested again when the pump is further increased, until the threshold for a double pulse is crossed. In this regime, there is sufficient energy to sustain two pulses in the cavity, and the pulse prefers to split. The onset of a double pulse is usually (but not always) observed as a fringe pattern on the pulse spectrum, as plotted in Figure~\ref{fig:MLbehavior}d, due to spectral interference between the two pulses \cite{spectralfringes}. Analysis of multi-pulse operation and single-pulse stabilization can be found in \cite{multipulse}.

\section{Beyond the Standard Cavity Design} \label{sc.noveldesigns}

So far, we described the operation concepts of the standard cavity design. The need to optimize the laser performance or to overcome inherent disadvantages and limitations motivated many extraordinary cavity designs. Since a comprehensive survey of all nonstandard cavity designs is beyond the scope of this work, \mbox{in the following}, we review our personal selection of published attempts to address these issues, some of our own making.

\subsection{Optimization of Laser Parameters: Compactness, Optical Elements, Stability and Pulse Duration}

\label{sc.optimization}

Various parameters of operation can be optimized using advanced cavity designs. Investigation of two- and three-mirror cavity configurations \cite{optimalres} demonstrated highly compact cavities, in which the Kerr efficiency (Equation~\eqref{eq:gamma}) was found to be maximal in a three-mirror configuration. Optimization guidelines for the crystal length were given in \cite{hardaperture} considering the estimated Rayleigh range of the laser mode in the crystal. Another novel cavity design was demonstrated in \cite{noveldesign1} by replacing the flat end mirror in the long arm with a curved one. The result is a fundamentally different diagram of the stability zones (compared to the conventional diagram shown in Figure~\ref{fig:modeonEM}) with a considerable improvement in the laser intensity stability. Pulse duration can also be minimized by maximizing the spectral broadening of the pulse (by SPM; see Section~\ref{sc:optkerreff}) inside the Ti:sapphire crystal. It was found that to maximize SPM, it is important to consider the physical arrangement of the elements in the cavity in order to achieve a symmetric dispersion distribution. When both cavity arms are compensated for dispersion independently \cite{octavespan}, the nonlinear response can be twice as strong as the standard dispersion compensation, since it maximizes the nonlinear response in both forward and backward propagation through the crystal. Thus, for maximal spectral broadening, it is beneficial to consider either a four-prism sequence in a ring cavity configuration or to include a prism pair/chirped mirrors in the short arm, also. Spectral broadening can also be achieved using an additional Kerr medium \cite{n2additional6}, thus enhancing the intra-cavity nonlinearity. This results in a mode-locked operation that can be sustained with lower pulse energy. This effect is of great importance for frequency comb-based measurements, where a higher repetition rate is preferred, which inherently requires a lower pulse energy. Usually, the additional Kerr medium is a Brewster-cut window placed in an additional focus in the cavity. However, due to the beam expansion in the tangential dimension upon refraction into the Brewster window, this configuration does not fully exploit the Kerr nonlinearity of the material. Recently, we presented a novel cavity design \cite{MLnoCW}, illustrated in Figure~\ref{fig:novelcavity}b, with an additional planar-cut (non-Brewster) and anti-reflection (AR) coated Kerr medium in normal incidence, which allows full exploitation of the nonlinear response, due to the tighter, astigmatic-free focusing compared to a Brewster-cut window. In addition, another interesting phenomenon was observed: with sufficient enhancement of the nonlinearity, the ML threshold can be lowered all the way down to the CW threshold, as illustrated in Figure~\ref{fig:novelcavity}d. When the two thresholds are equal, ML can be obtained directly from zero CW oscillation.


\subsection{Elimination of Nonlinear Astigmatism} \label{Sc.nononlinearastig}

\begin{figure}
	\centering
	\centerline{\includegraphics[page=12, clip, trim=1.3cm 0cm 1.3cm 0cm, width=8cm]{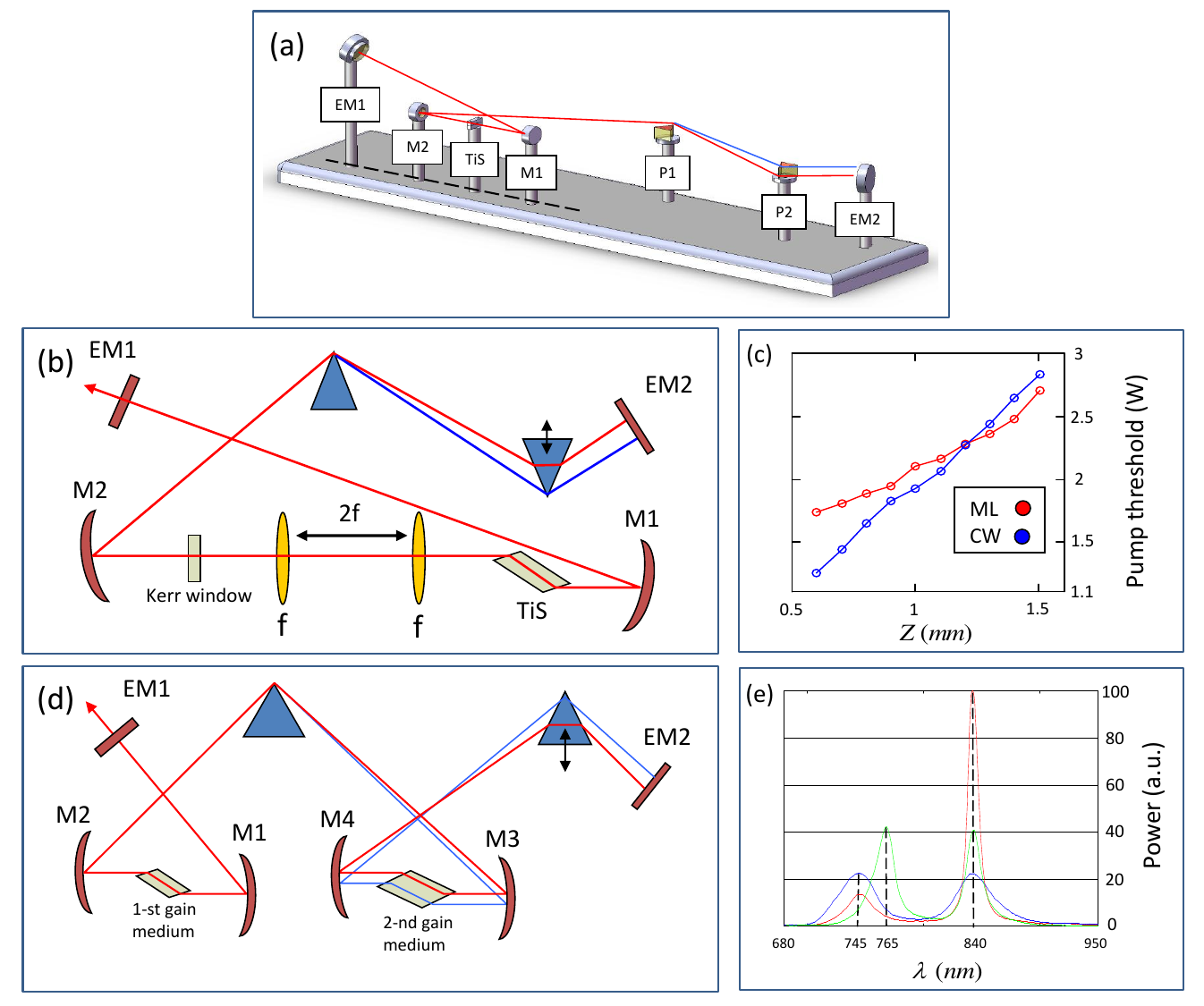}}
	\caption[Novel cavity configurations]
	{{\small{(\textbf{a}) Cavity configuration for canceling the nonlinear astigmatism. The long arm remains parallel to the main plane of the cavity, while the short arm is raised above the main plane, allowing for each of the curved mirrors to compensate for the astigmatism of the other and to introduce a planar-cut astigmatic-free Ti:sapphire crystal. (\textbf{b}) Cavity configuration for enhancement of the intra-cavity Kerr nonlinearity. An additional planar-cut Kerr window is inserted near the image point of the Ti:sapphire crystal, created by a lens-based $1\times 1$ telescope of focal length $f$. (\textbf{c}) The result for the configuration in (\textbf{b}): ML (red) and CW (blue) pump thresholds as a function of $Z=\delta-\delta_{1}$. The intra-cavity nonlinearity was enhanced by adding a $3$ mm-long window of borosilicate crown glass (BK7). At low values of $Z$, the ML threshold is higher than the CW threshold. By increasing $Z$, the ML threshold curve eventually crosses the CW threshold at \mbox{$Z_{c}\approx$ 1.2 mm}, where the intra-cavity CW power needed to initiate pulsed operation drops to zero. (\textbf{d}) The cavity configuration for intra-cavity control of spectral amplitude. A second Ti:sapphire gain medium is placed at the Fourier plane of a $1\times 1$ telescope (mirrors $M3$ and $M4$) placed between the prisms, resulting in an additional gain, which is effectively inhomogeneous, due to the spatial dispersion inside the medium, thereby eliminating mode competition in the gain medium. \mbox{(\textbf{e}) The results} for the configuration in (\textbf{d}): control of spectral power, width and center position of the two lobes; a two-lobed spectrum with equal intensities between the lobes (green curve) can be deformed in real time by increasing the width of each lobe (blue curve) or the spectral power ratio between the lobes (red curve). The center wavelength of each lobe can also be controlled by shifting the pump spot in the second crystal (demonstrated in both the blue and \mbox{red curves}).}}}
	\label{fig:novelcavity}
\end{figure}

The problem of nonlinear Kerr lens astigmatism, discussed in Section~\ref{sc.additionastig}, can be completely eliminated using the novel cavity folding demonstrated in \cite{nononlinastig}. This configuration allows for the introduction of a planar-cut (non-Brewster) and AR-coated crystal, where the beam enters the crystal at normal incidence and the spatial mode in both planes remains identical. Consequently, the crystal is astigmatic-free and does not introduce linear or nonlinear astigmatism, thus eliminating Kerr lens astigmatism from the source. The curved mirrors then compensate for the astigmatism of one for the other by using a novel cavity folding, as illustrated in Figure~\ref{fig:novelcavity}a. Mirror $M2$ folds the beam in the main plane of the cavity, whereas $M1$ folds the beam upwards. Thus, the sagittal and tangential components of $M1$ exchange roles with those of $M2$, leading to exact cancellation of the linear astigmatism of one mirror by the~other.

\subsection{Two-Color Pulses by Intra-Cavity Gain Shaping}

Many applications require shaping and controlling the spectrum of a KLM Ti:sapphire lasers. Specifically, a multi-lobed (in particular, dual-lobed) spectrum is useful for Raman spectroscopy \cite{antistokes}, Raman microscopy \cite{twophoton,sampleimaging} and direct frequency comb spectroscopy \cite{combspec}. Spectral shaping inside the optical cavity, which steers the laser oscillation towards the desired spectrum, is by far more profitable than spectral shaping outside the optical cavity, since the latter is inherently lossy in nature, due to the unavoidable filtering of undesired frequencies.

Previous attempts to obtain pulses with dual-lobed (two-color) spectra can be divided into two categories. The first is intra-cavity loss shaping, which can be achieved, for example, by a double slit after the prism pair \cite{doubleslit}, by a three prism configuration using a separate ``second'' prism for each lobe \cite{threeprisms} or by dual output coupling using custom-designed wavelength-dependent reflectivity curves of two output couplers \cite{twoOCs}. Unfortunately, loss shaping is very sensitive to mode competition in the laser gain medium, which inflicts a stringent requirement that the lobes be perfectly symmetrical around the gain peak (equal gain) in order to coexist. The second category overcomes mode competition either by actively synchronizing two independent sources \cite{phaselocking,OPOs} or by passive synchronization using a shared Kerr medium in both cavities \cite{synchronization,twobeams,fourprisms}. These methods suffer from the considerable added complexity of several independent oscillators, which require careful stabilization of timing jitter between the participating pulse trains.

We recently demonstrated the use of gain shaping, instead of loss shaping, in a single oscillator \cite{YEFETgainshaping}. \mbox{As illustrated in} Figure~\ref{fig:novelcavity}c, this can be achieved by placing another gain medium at a position in the cavity where the spectrum is spatially dispersed. In this additional gain medium, the gain profile can be tailored at will by controlling the spatial shape of the pump in this medium. Since the spectrum is spatially dispersed, mode competition is canceled in this additional gain medium (different frequencies do not share the same gain volume, rendering the gain effectively inhomogeneous). While placing a single gain medium in a spatially-dispersed position was considered in the past both \mbox{theoretically \cite{gainindisp1,gainindisp2}} and experimentally \cite{inhomogeneousgain}, however, such a laser is inherently pump-inefficient, since it requires pumping of a much wider volume in the gain medium well above medium. It is therefore most beneficial to combine a large homogeneous (efficient) gain with a small amount of inhomogeneous (inefficient) gain, where the small shaped inhomogeneous gain acts as a lever to steer the competition in the large homogeneous gain medium. Consequently, robust and complete control of the oscillation spectrum inside the optical cavity was achieved in a power preserving manner, as illustrated in Figure~\ref{fig:novelcavity}e.

\section{Summary}

The role that a Kerr lens mode-locked solid-state laser (and, especially, the Ti:sapphire laser) plays in the field of ultrafast physics cannot be underestimated. Since the Ti:sapphire laser is a common resident in most laboratories investigating ultrafast phenomena, we hope to have provided a detailed review that coherently summarizes, under a single notation, the important physical and experimental considerations involved in the design and realization of such an oscillator. The references given throughout this review provide information for additional and important topics, not included in this review, such as cavity alignment, self-starting and initiation of mode-locking, the optimal size of the hard aperture, oscillator noise, \textit{etc.} In conclusion, we aimed to illuminate the most important experimental considerations involved in the design of a Kerr lens mode-locked laser oscillator.



\section*{\noindent Acknowledgments}
\vspace{12pt}

This research was supported by the Israeli Science Foundation (grant \#807/09) and by the {Kahn~Foundation.}

\section*{\noindent Conflicts of Interest}
\vspace{12pt}

The authors declare no conflict of interest.

\appendix

\section*{\noindent Appendix}

\setcounter{table}{0}
\renewcommand\thetable{A\arabic{table}}

\section{Generalized Linear Astigmatism Compensation} \label{ap.astigmatism}

Equation \eqref{eq:astig} compensates for linear astigmatism only for the first stability limit, $\delta_{0}$. This is because the above expression for mirror astigmatism, $\Delta f(f,\theta)$, assumes a collimated beam in both arms of the cavity, which is only true at $\delta_{0}$, as seen in Figure~\ref{fig:modeonEM}. For any other stability limit, one must use a generalized expression for the curved mirror astigmatism that takes into account non-collimated beams, focused on the relevant end mirror. By considering the curved mirror as an imaging lens, one can calculate the distance of the point of the image relative to the curved mirror:

\begin{align}
 \label{eq:genastig}
 v_{s,t}(f,\theta,D)
 & =
 \frac{Df_{s,t}}{D-f_{s,t}}
\end{align} where $D$ represents the distance of the ``object''. For a collimated beam, $D=\infty$ and $v_{s,t}=f_{s,t}$. \mbox{For a non-collimated} beam, the ``object'' is effectively located at the end mirrors (see Figure~\ref{fig:geometricMODES}); hence, $D$ equals the length of the corresponding cavity arm. Using Equation~\eqref{eq:genastig}, one can compensate for linear astigmatism for each stability limit, $\delta_{i}$, separately, by solving:
\begin{align}
 \label{eq:fullastig}
 \Delta v(f_{1},\theta_{1},D1_{i})+\Delta v(f_{2},\theta_{2},D2_{i})=\Delta L
\end{align}

The corresponding values of $D1_{i}$ and $D2_{i}$ are given in Table~\ref{tab:astigarms}. As an example, we can examine typical cavity parameters: curved mirrors radius of curvature $R$ = 10 cm, short arm length $L_{1}$ = 30 cm, long arm length $L_{2}$ = 60 cm and crystal length $3$ mm. Compensating linear astigmatism for the first stability limit, $\delta_{0}$, results in $\theta_{1}=\theta_{2}=6.15^{0}$. In order to compensate for the second stability limit, $\delta_{1}$, the angle, $\theta_{2}$, must be reduced to $5.53^{0}$.

\begin{table}
	\centering \small
	\begin{tabular}{ccccccc} \toprule
		\boldmath{$\delta_{i}$} & \boldmath{$D1_{i}$} & \boldmath{$D2_{i}$}\\
		\midrule
		
		$\delta_{0}$ & $\infty$ & $\infty$\\
		$\delta_{1}$ & $\infty$ & $L_{2}$\\
		$\delta_{2}$ & $L_{1}$ & $\infty$\\
		$\delta_{3}$ & $L_{1}$ & $L_{2}$\\
		\bottomrule
	\end{tabular}
	\caption[Astigmatism compensation values]{{Astigmatism compensation values of the long and short cavity arms for each stability \mbox{limit, $\delta_{i}$.}}}\label{tab:astigarms}
\end{table}

Astigmatism compensation can be further generalized to compensate for $\delta$ values within the stability zones of the cavity (not necessarily at the stability limits) by using a complex representation of the imaging expression given in Equation~\eqref{eq:genastig}, which simulates a Gaussian mode (in contrast to the stability limits where the mode can be represented using geometrical optics, as illustrated in Figure~\ref{fig:geometricMODES}). However, it is easier to set the values of the folding angles, $\theta_{1}$ and $\theta_{2}$, to compensate for the nearest stability limit and then to fine-tune one of the folding angles until a circular mode is achieved.


\end{document}